\documentclass[12pt]{article}
\usepackage{wider}
\usepackage{xcolor}
\usepackage{soul}
\usepackage{comment}
\usepackage{cite}
\usepackage{fnpct}
\usepackage[multiple]{footmisc}

\usepackage{utfsym}

\begin{document}

\newcommand{\sqvb}{\ensuremath{ \langle \!\langle 0 |} }
\newcommand{\sqvk}{\ensuremath{ | 0 \rangle \!\rangle } }
\newcommand{\sqvn}{\ensuremath{ \langle \! \langle 0 |  0 \rangle \! \rangle} }
\newcommand{\wh}{\ensuremath{\widehat}}
\newcommand{\be}{\begin{equation}}
\newcommand{\ee}{\end{equation}}
\newcommand{\bea}{\begin{eqnarray}}
\newcommand{\eea}{\end{eqnarray}}
\newcommand{\ra}{\ensuremath{\rangle}}
\newcommand{\la}{\ensuremath{\langle}}
\newcommand{\rra}{\ensuremath{ \rangle \! \rangle }}
\newcommand{\lla}{\ensuremath{ \langle \! \langle }}
\newcommand{\str}{\rule[-.125cm]{0cm}{.5cm}}
\newcommand{\pr}{\ensuremath{^{\;\prime}}}
\newcommand{\ppr}{\ensuremath{^{\;\prime \prime}}}
\newcommand{\da}{\ensuremath{^\dag}}
\newcommand{\as}{^\ast}
\newcommand{\eps}{\ensuremath{\epsilon}}
\newcommand{\ve}{\ensuremath{\vec}}
\newcommand{\ka}{\kappa}
\newcommand{\non}{\ensuremath{\nonumber}}
\newcommand{\lf}{\ensuremath{\left}}
\newcommand{\rt}{\ensuremath{\right}}
\newcommand{\al}{\ensuremath{\alpha}}
\newcommand{\dfn}{\ensuremath{\equiv}}
\newcommand{\ga}{\ensuremath{\gamma}}
\newcommand{\ti}{\ensuremath{\tilde}}
\newcommand{\wti}{\ensuremath{\widetilde}}
\newcommand{\hs}{\ensuremath{\hspace*{.5cm}}}
\newcommand{\bet}{\ensuremath{\beta}}
\newcommand{\om}{\ensuremath{\omega}}
\newcommand{\kp}{\ensuremath{\kappa}}

\newcommand{\cO}{\ensuremath{{\cal O}}}
\newcommand{\cS}{\ensuremath{{\cal S}}}
\newcommand{\cF}{\ensuremath{{\cal F}}}
\newcommand{\cX}{\ensuremath{{\cal X}}}
\newcommand{\cZ}{\ensuremath{{\cal Z}}}
\newcommand{\cG}{\ensuremath{{\cal G}}}
\newcommand{\cR}{\ensuremath{{\cal R}}}
\newcommand{\cV}{\ensuremath{{\cal V}}}
\newcommand{\cC}{\ensuremath{{\cal C}}}
\newcommand{\cP}{\ensuremath{{\cal P}}}
\newcommand{\cH}{\ensuremath{{\cal H}}}
\newcommand{\cN}{\ensuremath{{\cal N}}}
\newcommand{\cE}{\ensuremath{{\cal E}}}

\newcommand{\pup}{\ensuremath{^{(p)}}}
\newcommand{\prpr}{\ensuremath{\prime \prime }}

\newcommand{\hsp}{\ensuremath{\hspace*{5mm} }}
\newcommand{\sbp}{\ensuremath{_{[p]} }}

\newcommand{\erf}{\mathop{\mathrm{erf}}}

\mathchardef\mhyphen="2D
\newcommand\utoe{\mathit{u{\mhyphen}\!e}}


\newcommand{\xxx}[1]{}
\newcommand{\yyy}[1]{}
\newcommand{\zzz}[1]{}
\newcommand{\xyz}[1]{}
\newcommand{\ttt}[1]{}
\newcommand{\tstart}[1]{}
\newcommand{\tstop}[1]{}
\newcommand{\tstartb}[1]{}
\newcommand{\tstopb}[1]{}

\title{\bf How can quantum field operators 
encode entanglement?}

%
%
%
%
%
%
%
%
%
%

\author{
Mark A. Rubin\\
\mbox{}\\
markallenrubin@yahoo.com\\
}
\date{\mbox{}}
\maketitle
\begin{abstract}
We present techniques to construct the Deutsch-Hayden representation for  quantum field operators and apply them to an entangled state of identical nonrelativistic spin-1/2 fermions localized in well-separated spatial regions.  Using these entangled field operators 
we construct operators measuring spin in localized spatial regions, and verify that matrix elements of the spin-measurement operators in the information-free Deutsch-Hayden state yield the expected correlations between pairs of both entangled and unentangled particles. The entangled Deutsch-Hayden-representation field operators furnish an explicitly separable description of the entangled system.

\mbox{}

\noindent Key words: entanglement, quantum field theory, Deutsch-Hayden representation, identical particles, separability, locality.

\end{abstract}

\section{Introduction}\label{SecIntro}
\xxx{SecIntro}

\tstartb{2:22PM Tues June 21 2022}

The phenomenon of entanglement is  
commonly regarded as demonstrating that quantum theory is nonlocal, with the example of Einstein-Podolsky-Rosen-Bohm (EPRB) correlations \cite{EPR},\cite[pp. 614-623]{Bohm1951} most often cited. 
In the EPRB scenario, perfect anticorrelation between
spin measurements on pairs of spatially-separated spin-1/2 particles in the singlet state, an entangled state, occurs when the measurements are made along the same axis for each of the paired particles.  From these anticorrelations and  the principle of locality --- ``all physical effects  are propagated with finite, subluminal velocities, so that no effects can be communicated between systems separated by a space-like interval'' 
\cite{Howard1985} --- the conclusion is drawn that information determining the outcomes of spin measurements along any axis that might be measured must reside in each particle (see, e.g.,  \cite{Bell64,Bell81,Farris95,Norsen17}), information  Mermin has termed ``instruction sets" \cite{Mermin81}.

When, however, the results of 
spin measurements on different axes for each particle are also taken into account,   Bell's theorem \cite{Bell64} shows that the correlations observed between the results measured on members of singlet-state pairs are in fact  {\em inconsistent} with the existence of such instruction sets. 

From this contradiction between locality and the experimentally-verified (see, e.g., \cite{Rauch18} and references therein to earlier experiments) predictions of quantum mechanics the conclusion is generally drawn that the principle of locality does not in fact hold in the physical world (see e.g., \cite{Bell81,Bell90}). Norsen, for example, states that  ``what should be concluded from the experimentally observed violations of Bell-type inequalities is [that we must]
simply conclude that locality -- that the prohibition on faster-than-light causation that seems somehow to be implied by relativity theory -- is false. Relativistic local causality is wrong, is in conflict with experimental data. Faster-than-light causal influences
really exist in Nature!'' \cite[p. 234]{Norsen17}.  \tstartb{1:23PM Wed June 22, 2022} 

Some authors argue that locality can be preserved if one relinquishes instead 
the principle of separability ---``any two spatially separated systems possess their own separate real states'' \cite{Howard1985}  ---  so that information in some sense resides holistically in both particles. Howard, for example, claims  that ``these [Bell] 
experiments should be interpreted as refuting the separability principle'' \cite{Howard1985}.  Brown and Timpson  argue that ``how  a  non-separable  theory  can  locally explain Bell-inequality violating correlations would be that the correlations are entailed by some suitable  non-separable joint state'' \cite{BrownTimpson14}.

We are not compelled to accept these conclusions regarding the nonlocality or the nonseparability of quantum mechanics 
if we consider quantum mechanics in the  Everett interpretation.  Bell's theorem does not apply to quantum mechanics in the  Everett interpretation (see e.g., \cite{Eberhard78,Page82,Stapp85,Vaidman90,Price95,Tipler00,DeutschHayden00,Rubin01,Vaidman02,Bacciagaluppi02,TimpsonBrown02,HewittHorsman09,Rubin11,Deutsch12,Vaidman15}). 
As Vaidman succinctly puts it, ``Bell's theorem \ldots cannot get off the ground in the framework of the [Everett interpretation] because it requires a single outcome of a quantum experiment''\cite{Vaidman02}.

Still, we would like to know: If not as instruction sets, 
then {\em how}\/ is  the information that leads to the correlations between entangled particles, both perfect correlations in the same-axis case and imperfect correlations in the different-axis case,  encoded in the  quantum formalism? 

In the present paper we focus on the case of entangled identical particles in nonrelativistic quantum field theory. We consider 
a state containing three spin-1/2 fermions, two of which are entangled. Correlations between spin measurements on the two entangled particles will be different from those between either of the entangled particles and the third unentangled particle. What are the elements of the quantum-field-theoretic formalism responsible for this difference? How is this difference consistent with the identity of the particles, each of which is but an excitation of the underlying quantum field?    Does the information that determines the spin correlations divide up into portions that can be thought of as inhering respectively in individual particles, thus satisfying the principle of separability?

The answers to these questions begin with the observation that to be able to specify that two of the three particles are entangled and the other is not, 
 all the particles must have distinct spatial locations to at least some degree.  One cannot  speak of ``particle 1'' or ``particle 2''  or ``particle 3,'' but, if the particles are sufficiently well-localized  in distinct spatial regions,  
 one can speak of ``this particle over here'' or ``that particle over there.''  It is thus natural to look to the field operators as elements of the formalism in which information pertaining to each particle could be encoded, since the field operators are indexed with spatial location from the outset.

However, although the field operators have spatial locations, in the usual representation of 
quantum field theory they do not by themselves encode information about entanglement or anything else. 
``The
operator corresponding to a given observable represents not the value of the observable, but rather all the values that the observable can assume under various conditions, the values themselves being the eigenvalues.\ldots The dynamical variables of a system, being operators, do no represent the system other than generically.  That is, they represent not the system as it really is, but rather all the situations in which the system might conceivably find itself.\ldots
Which situation a system is actually in is specified by the state vector.  
Reality is therefore described jointly by the dynamical variables and the state vector''\cite[p. 182]{DeWitt72}.
 To have field operators that in and of themselves encode information, in particular but not limited to information about spin correlations, we must make use of the Deutsch-Hayden representation.

 \tstartb{9:26am, Fri. June 24 2022}
 \tstopb{~10:45am, Fri. June 24 2022 -- work ~ 1h. 15m.}
 \tstartb{11:40am, Fri. June 24 2022 -- make lunch, email Al -- ~1h}
 \tstopb{~12:15pm, Fri. June 24 2022 -- work ~35m.}
 \tstartb{1:02pm, Fri. June 24 2022 -- lunch ~45m.}

    In a seminal paper, Deutsch and Hayden \cite{DeutschHayden00}  show that quantum computational networks encode and transport information in a local manner.\footnote{Deutsch and Hayden\cite{DeutschHayden00} and Deutsch\cite{Deutsch12} argue that this proof of locality extends from quantum computational networks to all quantum systems as a consequence of the universality of quantum computation\cite{Deutsch89}  and the Beckenstein bound\cite{Beckenstein81}.  Marletto, Tibau Vidal and Vedral\cite{Marlettoetal21} question the
    applicability of this argument to fermions. See also Sec. \ref{SecDisc}  below.} 
A key step in demonstrating this is to point out that one can perform a unitary transformation to a representation in which the state vector is mapped to a standard state containing no physical information, while the information which is usually encoded in the state vector is transferred  to operators.    We refer to this representation as the Deutsch-Hayden representation, and to the transformation from the usual  representation to the Deutsch-Hayden representation as the Deutsch-Hayden transformation.\footnote{In a previous paper\cite{Rubin11}  I referred to the Deutsch-Hayden ``picture.'' I now feel that this is inappropriate terminology.  The Schr\"{o}dinger and Heisenberg pictures are different ways of describing time evolution of quantum systems. The Deutsch-Hayden transformation maps all information residing in the state vector at a given time to the operators. If time evolution is  subsequently performed using the Heisenberg picture, the state vector will remain the information-free standard state, since the state vector in the Heisenberg picture is constant. 
So, it is more natural to perform time evolution in the Deutsch-Hayden representation using the Heisenberg picture.} 
\tstopb{3:34pm, Fri. June 24 2022  -- work 2h. 3m.}

To see in detail how information regarding entanglement is encoded in quantum-field-theoretic operators, we will perform a Deutsch-Hayden transformation from an entangled quantum field state and obtain the field operators in the Deutsch-Hayden representation.  
Our approach to explicitly performing this Deutsch-Hayden transformation and verifying its properties utilizes four key ingredients:
\begin{enumerate}
\item Auxiliary fields (Sec.~\ref{SecPhysauxfieldops}). These additional fermionic fields, one for each particle in the system, carry no physical information but allow for the effective locality of the Deutsch-Hayden transformation of the field operators (Sec. \ref{SecEffloc}), as first pointed out in \cite{Rubin11}. 
\item Widely-separated wavepackets (Sec.~\ref{SecWSW}), so that we can distinguish otherwise-identical particles by their approximate spatial location. 
\item Two-step procedure (Sec.~\ref{SecTwostep}) for obtaining the Deutsch-Hayden transformation appropriate to an entangled state using the simpler Deutsch-Hayden transformation appropriate to a ``nearby''  unentangled state.
\item Localized spin operators (Sec.~\ref{SecLocspinops}) constructed using aperture functions matched to the widely-separated wavepackets,  to measure the spin associated with each particle. \end{enumerate}

This paper is organized as follows.  In Sec.~\ref{SecunentDHops0} we define an unentangled state containing three identical nonrelativistic spin-1/2 fermions localized in three different well-separated regions, with the particle localized in region 1 spin-up and the particles respectively localized in regions 2 and 3 spin-down.  We define operators corresponding to the measurement of spin in a specified localized region,  verify that the unentangled three-particle state is an eigenstate of these operators,  and verify that it has the expected pairwise spin correlations. In Sec. \ref{SecDHrep} we construct a Deutsch-Hayden transformation for this state, compute the Deutsch-Hayden-transformed field operators, and point out the sense in which, and the conditions under which, this transformation is effectively local.   
In Sec.~\ref{SecEntDHrep} we present the two-step procedure for obtaining a Deutsch-Hayden transformation for an  entangled state from the Deutsch-Hayden representation for an unentangled state from which it can be generated, present  an operator that generates the entangled state, compute the entangled Deutsch-Hayden field operators and localized spin operators, and verify that the latter have the correct expectation values and correlations in the information-free Deutsch-Hayden state.  We present and discuss our conclusions in 
Sec.~\ref{SecDisc}.  In  Appendix~A 
we calculate spin expectation values and correlations for the corresponding first-quantized system. 
Appendix~B 
examines the role played by auxiliary fields in the formalism.

\tstart{Thurs. June 2 12:00pm}

\section{Unentangled state in the usual representation}\label{SecunentDHops0}
\xxx{SecunentDHops0}

\subsection{Physical and auxiliary field operators}\label{SecPhysauxfieldops}

\xxx{SecPhysauxfieldops}

We will employ the term ``usual representation'' to refer to states and  operators on which a Deutsch-Hayden transformation has not been performed.  

We will be working at a single time that  we will take to be the 
time $t=0$\/ at which Heisenberg-picture states and operators are equal to their Schr\"{o}dinger-picture counterparts.  
The particles in the system are nonrelativistic spin-1/2 fermions; we will associate an index 1 to a particle which is
spin-up with respect to the $x_3$\/ axis, and an index 2 to a particle which is spin-down with respect to this axis. So, a particle with spin-up ($i=1$\/) or spin-down ($i=2$\/) is created 
at point $\vec{x}$\/ at time  $t=0$\/ by the creation operator
$\wh{\phi}_i^\dag(\vec{x})$\/ satisfying the anticommutation relations
\be
\{\wh{\phi}_i(\vec{x}), \wh{\phi}_j^\dag(\vec{y})\}=\delta^3(\vec{x}-\vec{y})\delta_{i,j}, \hspace*{.5cm} i,j=1,2,\label{ETAR1}
\ee
\xxx{ETAR1}
\be
\{\wh{\phi}_i(\vec{x}), \wh{\phi}_j(\vec{y})\}=\{\wh{\phi}_i^\dag(\vec{x}), \wh{\phi}_j^\dag(\vec{y})\}=0,\hspace*{.5cm} i,j=1,2,\label{ETAR2}
\ee
\xxx{ETAR2}
as well as
\be
\wh{\phi}_i(\vec{x})|0\ra=0,\hspace*{.5cm}i=1,2, \label{killvac}
\ee
\xxx{killvac}
where $|0\ra$\/ is the usual vacuum state. 

In addition to these familiar creation and annihilation field operators, which we will refer to as physical field operators, we introduce auxiliary field operators 
$\wh{\alpha}_{(i)}^\dag(\vec{x})$\/, $i=1,2,3$\/, satisfying 
\be
\{\wh{\alpha}_{(i)}(\vec{x}), \wh{\alpha}_{(j)}^\dag(\vec{y})\}=\delta^3(\vec{x}-\vec{y})\delta_{i,j},, \hspace*{.5 cm} i,j=1,2,3,\label{auxETAR1}
\ee
\xxx{auxETAR1}
\be
\{\wh{\alpha}_{(i)}(\vec{x}), \wh{\alpha}_{(j)}(\vec{y})\}=\{\wh{\alpha}_{(i)}^\dag(\vec{x}), \wh{\alpha}_{(j)}^\dag(\vec{y})\}=0,\hspace*{.5 cm} i,j=1,2,3,\label{auxETAR1b}
\ee
\xxx{auxETAR1b}
as well as
\be
\wh{\alpha}_{(i)}(\vec{x})|0\ra=0,\hspace*{1cm}i=1,2,3, \label{auxkillvac}
\ee
\xxx{auxkillvac}
and anticomuting with the physical field operators:
$$
\{\wh{\phi}_i(\vec{x}), \wh{\alpha}_{(j)}(\vec{y})\}=\{\wh{\phi}_i^\dag(\vec{x}), \wh{\alpha}_{(j)}(\vec{y})\}=
\{\wh{\phi}_i(\vec{x}), \wh{\alpha}_{(j)}^\dag(\vec{y})\}=\{\wh{\phi}_i^\dag(\vec{x}), \wh{\alpha}_{(j)}^\dag(\vec{y})\}
=0,
$$
\be
\hspace*{8cm}i=1,2, \hspace*{.5cm}j=1,2,3.\label{auxETAR2}
\ee
\xxx{auxETAR2}
We will refer to (\ref{ETAR1})-(\ref{auxETAR2}), including  (\ref{killvac}) and (\ref{auxkillvac}), as the equal-time anticommutation relations (ETARs).

\tstop{Thurs June 2, 12:13pm}
\tstart{12:28pm Thurs 6/2/22}

\subsection{Widely-separated wavepackets}\label{SecWSW}

\xxx{SecWSW}

To construct the initial state we will make use of three complex c-number functions, $\psi_{\{i\}}(\vec{x})$\/, $i=1,2,3$\/, which
are normalized,
\be
\int d^3\vec{x} |\psi_{\{i\}}(\vec{x})|^2=1,\hspace*{.5cm }i=1,2,3,\label{wwnormalization}
\ee
\xxx{wwnormalization}
and which correspond 
to the three regions in which the particles are localized.  We will refer to these as physical  wavefunctions.  The key property that these functions possess is that 
they are  ``widely-separated wavepackets,'' i.e., their supports are approximately nonoverlapping:
\be
\psi_{\{i\}}(\vec{x})\psi_{\{j\}}(\vec{x}) \approx 0, \hspace*{.5 cm} i\neq j,\hspace*{.5cm } i=1,2,3.\label{wswconditions}\ee
\xxx{wswconditions}
\yyy{OD-500 b3 \textcolor{green}{\usym{1F5F8}} }
We will refer to 
 (\ref{wswconditions}) as the  widely-separated wavepacket (WSW) conditions,  and in what follows we will  treat them as exact   equalities. 
\tstart{2:50pm Thurs June 2, 2:51pm}

\subsection{Unentangled state}\label{SecUnentState}

\xxx{SecUnentState}

For the unentangled three-particle state, in the usual representation, we take
$$
\hspace*{-2 in}|\psi_{1,\{1\};2,\{2\};2,\{3\}}\ra=\int   d^3\vec{x}_1 d^3\vec{x}_2 d^3\vec{x}_3 
                                                        d^3\vec{y}_1 d^3\vec{y}_2 d^3\vec{y}_3  
$$
$$
\psi_{\{1\}}(\vec{x}_1)\psi_{\{2\}}(\vec{x}_2)\psi_{\{3\}}(\vec{x}_3) 
\psi_{(1)}(\vec{y}_1)\psi_{(2)}(\vec{y}_2)\psi_{(3)}(\vec{y}_3)
$$
\be
\hspace*{2 in}\wh{\phi}_1^\dag (\vec{x}_1)\wh{\phi}_2^\dag (\vec{x}_2)\wh{\phi}_2^\dag (\vec{x}_3)
\wh{\alpha}_{(1)}^\dag (\vec{y}_1)\wh{\alpha}_{(2)}^\dag (\vec{y}_2)
\wh{\alpha}_{(3)}^\dag (\vec{y}_3)|0\ra.\label{unentusualstate}   
\ee
\xxx{unentusualstate}
\yyy{OD-440 b1, OD-452 b-1 \textcolor{green}{\usym{1F5F8}}}
The auxiliary-field wavefunctions, $\psi_{(i)}(\vec{x})$\/, $i=1,2,3,$\/
are complex c-number functions that  are completely arbitrary except for normalization:
\be
\int d^3\vec{x} |\psi_{(i)}(\vec{x})|^2=1,\hspace*{.5cm }i=1,2,3.\label{auxnormalization}
\ee
\xxx{auxnormalization}

%
In the usual representation, the auxiliary fields  do not appear in operators corresponding to measurement of any physical quantity.  Since we are always free to work in the usual representation, expectation values in the state 
(\ref{unentusualstate}) of operators that are functions only of physical field operators will be independent of the arbitrary but (importantly) normalized auxiliary-field wavefunctions.  This will be true at all times,  since  auxiliary fields do not appear in the Hamiltonian, and the Hamiltonian only contains even powers of fermionic fields,   
   so they do not evolve in time.   For the same reasons it will also be true of expectation values in the entangled state generated from the unentangled state using the operator $\wh{H}$\/ introduced in Sec. \ref{SecGenent}
, in whatever representation they are computed, as will be seen in the computations presented below.
%
%

The physical predictions of the theory thus depend only on the physical operators and wavefunctions   
  and would be unchanged if the auxiliary operators and wavefunctions were removed. However, as was first pointed out in \cite{Rubin11} and as will be discussed  below in Sec.~\ref{SecEffloc}, Sec.~\ref{SecDisc}, and Appendix~B, the inclusion of the auxiliary fields allows for the construction of an effectively local Deutsch-Hayden transformation.


\subsection{Localized spin operators}\label{SecLocspinops}

\xxx{SecLocspinops}

We now wish to that the state (\ref{unentusualstate}) in fact represents three unentangled particles localized in regions 1, 2, and 3, where region $i$\/ is that volume where the support of $\psi_{\{i\}}(\vec{x})$\/ is non-negligible, and where the particle in region 1 is spin-up and the particles in regions 2 and 3 are spin-down. 
To this end we define operators corresponding to the measurement of spin, in units of $\hbar/2$\/, in a localized region along a specified axis.  

The operator measuring, at point $\vec{x}$\/,  the density of spin in the direction of a unit vector $\vec{u}$\/   is
\be
\wh{\cS}_{\vec{u}}(\vec{x}) = \wh{\cN}_{\vec{u},1}(\vec{x}) - \wh{\cN}_{\vec{u},2}(\vec{x}),\label{spindensity}
\ee
\xxx{spindensity}
\yyy{inferred from OD-235 b1  \textcolor{green}{\usym{1F5F8}}}
where 
\be
\wh{\cN}_{\vec{u},i}(\vec{x})=\wh{\phi}_{\vec{u},i}^\dag(\vec{x})\wh{\phi}_{\vec{u},i}(\vec{x}),\hspace*{.5in}i=1,2
\label{numberdensity}
\ee
\xxx{numberdensity}
\yyy{inferred from OD-235 b2  \textcolor{green}{\usym{1F5F8}}}
are the number densities for particles which are spin-up ($i=1$\/) and spin-down ($i=2$\/) along $\vec{u}$\/, and where, referring to  \cite[eqs. (A2a),(A2b)]{Greenbergeretal90}, the field operators for spin defined along $\vec{u}$\/ ($\wh{\phi}_{\vec{u},i}$, 
$i=1,2$\/) are related to those for spin along  $x_3$\/ ($\wh{\phi}_i$\/, $i=1, 2$\/)  by the relations 
\bea
\wh{\phi}_{\vec{u},1}^\dag(\vec{x})&=&e^{-i\phi/2}\cos(\theta/2)\wh{\phi}_1^\dag(\vec{x})+e^{i\phi/2}\sin(\theta/2)\wh{\phi}_2^\dag(\vec{x}),\label{phidagu1}\\
\wh{\phi}_{\vec{u},2}^\dag(\vec{x})&=&-e^{-i\phi/2}\sin(\theta/2)\wh{\phi}_1^\dag(\vec{x})+e^{i\phi/2}\cos(\theta/2)\wh{\phi}_2^\dag(\vec{x})\label{phidagu2}\eea
\xxx{phidagu1,phidagu2}
\yyy{OD-235 b6, OD-237 b1 \textcolor{green}{\usym{1F5F8}}}
and their adjoints, with
\be
\vec{u}=(\sin \theta \cos \phi, \sin \theta \sin \phi, \cos \theta). \label{unitvector}
\ee
\xxx{unitvector}
\yyy{OD-565 b-1 \textcolor{green}{\usym{1F5F8}}}

Next, define three ``aperture functions,''\footnote{So named to suggest the aperture of a physical device that measures the spin in a region, say by detecting the associated magnetic field of the particles.}  each corresponding to one of the regions in which the particles are localized. These functions take on at each point $\vec{x}$\/ either the value 0 or the value 1 and have nonoverlapping support:
\be
A_{\{i\}}(\vec{x})A_{\{j\}}(\vec{x})=\delta_{i,j}A_{\{j\}}(\vec{x}),\hspace*{.5 in}
i,j=1,2,3.\label{Asquared}
\ee
\xxx{Asquared}
\yyy{OD-238 b4,5  \textcolor{green}{\usym{1F5F8}}}
Each aperture function is matched to a physical wavepacket, in that it has value 1 where the wavepacket has nonnegligible value and 0 elsewhere, so 
\be
 A_{\{i\}}(\vec{x})\psi_{\{j\}}(\vec{x})\approx\delta_{i,j}\psi_{\{j\}}(\vec{x}),\hspace*{.5 in} i,j=1,2,3. \label{Apsimatch}
 \ee
 \xxx{Apsimatch}
 \yyy{OD-235 b3,4 \textcolor{green}{\usym{1F5F8}}}

\tstopb{11:59pm Wed., June 29, 2022 --- work 59m.}

 We will treat (\ref{Apsimatch}) as an exact equality. 
 Note that, with the normalization conditions (\ref{wwnormalization}),  (\ref{Apsimatch}) implies
 \be
 \int d^3\vec{x} A_i(\vec{x}) |\psi_j(\vec{x})|^2=\delta_{i,j},\hspace*{.5 in} i,j=1,2,3.\label{apertureint}
 \ee
 \xxx{apertureint}
 We will refer to (\ref{Asquared})-(\ref{apertureint}) as the aperture conditions.
  
  Using the aperture functions, define the operators measuring spin along $\vec{u}_{\{i\}}$\/ in region~$i$\/:
 \be
 \wh{S}_{\vec{u}_{\{i\}},\{i\}}=\wh{N}_{\vec{u}_{\{i\}},1,\{i\}}-\wh{N}_{\vec{u}_{\{i\}},2,\{i\}},\hspace*{.5cm}i=1,2,3,\label{locspinopu}
  \ee
  \xxx{locspinopu}
  \yyy{OD-235 b1 w n changed to u \textcolor{green}{\usym{1F5F8}}}
  where
 \be
\vec{u}_{\{i\}}=(\sin \theta_{\{i\}} \cos \phi_{\{i\}}, \sin \theta_{\{i\}} \sin \phi_{\{i\}}, \cos \theta_{\{i\}}), \hspace*{.5cm} i=1,2,3.\label{unitvectori}
\ee
\xxx{unitvectori}
\yyy{OD-565 b-1 \textcolor{green}{\usym{1F5F8}}}  
\be
  \wh{N}_{\vec{u}_{\{i\}},j,\{i\}}=\int d^3 \vec{x} \; A_{\{i\}}(\vec{x}) \; \wh{\cN}_{\vec{u},j}(\vec{x}),
  \hspace*{.5cm}j=1,2,\hspace*{.5cm} i=1,2,3.
  \label{locnumopju}
  \ee 
  \xxx{locnumopju}
  \yyy{OD-235 b2 \textcolor{green}{\usym{1F5F8}}}
  Using 
 (\ref{numberdensity})-(\ref{phidagu2}) and  (\ref{locspinopu})-(\ref{locnumopju}),
  \bea
  \wh{S}_{\vec{u}_{\{i\}},\{i\}}=\int d^3 \vec{x} \; A_{\{i\}}(\vec{x}) 
  & & \! \! \! \! \! \! \! \! \!  \left[ \rule{0 mm}{5 mm} \cos \theta_{\{i\}}  \left(\wh{\phi}^\dag _1(\vec{x})\wh{\phi} _1(\vec{x})
                                 -\wh{\phi}^\dag _2(\vec{x})\wh{\phi} _2(\vec{x})\right) \right. \nonumber \\
  &  &\! \! \! \! \! \!+ \left.  \rule{0 mm}{5 mm}  \sin \theta_{\{i\}} \left(e^{i\phi_{\{i\}}} \; \wh{\phi}^\dag _2(\vec{x})\wh{\phi} _1(\vec{x})
                                 +e^{-i\phi_{\{i\}}}\; \wh{\phi}^\dag _1(\vec{x})\wh{\phi} _2(\vec{x})\right) \right].  \label{locspinopuexpl}  
  \eea
  \xxx{locspinopuexpl}
  \yyy{OD-238 b1 \textcolor{green}{\usym{1F5F8}} }

  Using the ETARs (\ref{ETAR1})-(\ref{auxETAR2}),   (\ref{unentusualstate}), and the aperture conditions (\ref{Asquared})-(\ref{apertureint}) 
   with (\ref{locspinopuexpl}),
  \bea
   \wh{S}_{\vec{u}_{\{1\}},\{1\}}|\psi_{1,\{1\};2,\{2\};2,\{3\}}\ra &=& 
   \cos \theta_{\{1\}}|\psi_{1,\{1\};2,\{2\};2,\{3\}}\ra \nonumber \\
   &&+ \sin \theta_{\{1\}} e^{i\phi_{\{1\}}}  |\psi_{2,\{1\};2,\{2\};2,\{3\}}\ra \label{S1psiunent} \\ 
      \wh{S}_{\vec{u}_{\{2\}},\{2\}}|\psi_{1,\{1\};2,\{2\};2,\{3\}}\ra &=& 
  -\cos \theta_{\{2\}}|\psi_{1,\{1\};2,\{2\};2,\{3\}}\ra \nonumber \\ 
  &&+ \sin \theta_{\{2\}} e^{-i\phi_{\{2\}}}  |\psi_{1,\{1\};1,\{2\};2,\{3\}}\ra \label{S2psiunent}\\
  \wh{S}_{\vec{u}_{\{3\}},\{3\}}|\psi_{1,\{1\};2,\{2\};2,\{3\}}\ra&=&
  -\cos \theta_{\{3\}}|\psi_{1,\{1\};2,\{2\};2,\{3\}}\ra \nonumber \\
  && + \sin \theta_{\{3\}} e^{-i\phi_{\{3\}}}  |\psi_{1,\{1\};2,\{2\};1,\{3\}}\ra  \label{S3psiunent}
  \eea
  \xxx{S1psiunent,S2psiunent,S3psiunent  }   
  \yyy{OD-575 b1,b2,b3  \textcolor{green}{\usym{1F5F8}}    }  
where
$$
\hspace*{-2 in}|\psi_{2,\{1\};2,\{2\};2,\{3\}}\ra=\int   d^3\vec{x}_1 d^3\vec{x}_2 d^3\vec{x}_3 
                                                        d^3\vec{y}_1 d^3\vec{y}_2 d^3\vec{y}_3  
$$
$$
\psi_{\{1\}}(\vec{x}_1)\psi_{\{2\}}(\vec{x}_2)\psi_{\{3\}}(\vec{x}_3) 
\psi_{(1)}(\vec{y}_1)\psi_{(2)}(\vec{y}_2)\psi_{(3)}(\vec{y}_3)
$$
\be
\hspace*{2 in}\wh{\phi}_2^\dag (\vec{x}_1)\wh{\phi}_2^\dag (\vec{x}_2)\wh{\phi}_2^\dag (\vec{x}_3)
\wh{\alpha}_{(1)}^\dag (\vec{y}_1)\wh{\alpha}_{(2)}^\dag (\vec{y}_2)
\wh{\alpha}_{(3)}^\dag (\vec{y}_3)|0\ra\label{unentusualstate2}   
\ee
\xxx{unentusualstate2}
\yyy{OD-573 b-3  \textcolor{green}{\usym{1F5F8}}  }
$$
\hspace*{-2 in}|\psi_{1,\{1\};1,\{2\};2,\{3\}}\ra=\int   d^3\vec{x}_1 d^3\vec{x}_2 d^3\vec{x}_3 
                                                        d^3\vec{y}_1 d^3\vec{y}_2 d^3\vec{y}_3  
$$
$$
\psi_{\{1\}}(\vec{x}_1)\psi_{\{2\}}(\vec{x}_2)\psi_{\{3\}}(\vec{x}_3) 
\psi_{(1)}(\vec{y}_1)\psi_{(2)}(\vec{y}_2)\psi_{(3)}(\vec{y}_3)
$$
\be
\hspace*{2 in}\wh{\phi}_1^\dag (\vec{x}_1)\wh{\phi}_1^\dag (\vec{x}_2)\wh{\phi}_2^\dag (\vec{x}_3)
\wh{\alpha}_{(1)}^\dag (\vec{y}_1)\wh{\alpha}_{(2)}^\dag (\vec{y}_2)
\wh{\alpha}_{(3)}^\dag (\vec{y}_3)|0\ra\label{unentusualstate3}   
\ee
\xxx{unentusualstate3}
\yyy{OD-574 b 3 \textcolor{green}{\usym{1F5F8}}  }
$$
\hspace*{-2 in}|\psi_{1,\{1\};2,\{2\};1,\{3\}}\ra=\int   d^3\vec{x}_1 d^3\vec{x}_2 d^3\vec{x}_3 
                                                        d^3\vec{y}_1 d^3\vec{y}_2 d^3\vec{y}_3  
$$
$$
\psi_{\{1\}}(\vec{x}_1)\psi_{\{2\}}(\vec{x}_2)\psi_{\{3\}}(\vec{x}_3) 
\psi_{(1)}(\vec{y}_1)\psi_{(2)}(\vec{y}_2)\psi_{(3)}(\vec{y}_3)
$$
\be
\hspace*{2 in}\wh{\phi}_1^\dag (\vec{x}_1)\wh{\phi}_2^\dag (\vec{x}_2)\wh{\phi}_1^\dag (\vec{x}_3)
\wh{\alpha}_{(1)}^\dag (\vec{y}_1)\wh{\alpha}_{(2)}^\dag (\vec{y}_2)
\wh{\alpha}_{(3)}^\dag (\vec{y}_3)|0\ra\label{unentusualstate4}   
\ee
\xxx{unentusualstate4}
\yyy{OD-574 b-1  \textcolor{green}{\usym{1F5F8}}   }
 %
 
 \subsection{Eigenvalue-eigenvector link} \label{Seceelink}
 \xxx{Seceelink}
 
 If in each region we utilize localized spin operators measuring spin along the $x_3$\/ axis, i.e., we take
 \be
 \theta_{\{1\}}= \theta_{\{2\}}= \theta_{\{3\}}=0,\label{allthetas0}
 \ee
 \xxx{allthetas0}
 then, defining
 \be
 \wh{x}_3=(0,0,1)  \label{x3def}
 \ee
 and using (\ref{unitvectori}) and  (\ref{S1psiunent})-(\ref{S3psiunent}), we obtain
 \bea
   \wh{S}_{\wh{x}_3,\{1\}}|\psi_{1,\{1\};2,\{2\};2,\{3\}}\ra &=& 
   |\psi_{1,\{1\};2,\{2\};2,\{3\}}\ra  \label{S1zpsiunent} \\
   \wh{S}_{\wh{x}_3,\{2\}}|\psi_{1,\{1\};2,\{2\};2,\{3\}}\ra &=& 
  -|\psi_{1,\{1\};2,\{2\};2,\{3\}}\ra\label{S2zpsiunent}\\
  \wh{S}_{\wh{x}_3,\{3\}}|\psi_{1,\{1\};2,\{2\};2,\{3\}}\ra&=&
  -|\psi_{1,\{1\};2,\{2\};2,\{3\}}\ra  \label{S3zpsiunent}
  \eea
 \xxx{S1zpsiunent,S2zpsiunent,S3zpsiunent \textcolor{green}{\usym{1F5F8}}  }   
 \yyy{OD-575 b-3,b-2,b-1}
 That is, the unentangled state $|\psi_{1,\{1\};2,\{2\};2,\{3\}}\ra$\/ is a spin-up eigenstate of the operator measuring spin along the $x_3$\/ axis in regions 1, and a spin-down eigenstate of the operators measuring spin along the $x_3$\/ axis in regions 2 and 3.   
 
 \subsection{Pairwise spin correlations}\label{Secspincorr}
 \xxx{Secspincorr}
 
For arbitrary directions $\vec{u}_{\{1\}}$\/, $\vec{u}_{\{2\}}$\/,
$ \vec{u}_{\{3\}}$\/, the pairwise correlations between spins in different regions in the state 
$|\psi_{1,\{1\};2,\{2\};2,\{3\}}\ra$\/ can be computed using (\ref{S1psiunent})--(\ref{S3psiunent}).  From (\ref{locspinopuexpl}) we see that $\wh{S}_{\vec{u},\{i\}}$\/ is Hermitian, and from  the ETARs (\ref{ETAR1})-(\ref{auxETAR2}), the WSW conditions (\ref{wswconditions}) and the normalizations (\ref{wwnormalization}), (\ref{auxnormalization}) it follows that the four vectors
$|\psi_{1,\{1\};2,\{2\};2,\{3\}}\ra$\/, $|\psi_{2,\{1\};2,\{2\};2,\{3\}}\ra$\/, $|\psi_{1,\{1\};1,\{2\};2,\{3\}}\ra$\/ and
$|\psi_{1,\{1\};2,\{2\};1,\{3\}}\ra$\/ are normalized and mutually orthogonal.  Using in addition the aperture conditions (\ref{Asquared})-(\ref{apertureint}) we obtain \tstartb{10:36am, Tues June 28, 2022}
\tstopb{11:19am, Tues June 28, 2022 -- work 43m.}
\bea
\la\psi_{1,\{1\};2,\{2\};2,\{3\}}| \wh{S}_{\vec{u}_{\{1\}},\{1\}} \wh{S}_{\vec{u}_{\{2\}},\{2\}}  |\psi_{1,\{1\};2,\{2\};2,\{3\}}\ra&=&
-u_{\{1\},3}u_{\{2\},3} \label{spincorrunent12} \\ 
\la\psi_{1,\{1\};2,\{2\};2,\{3\}}| \wh{S}_{\vec{u}_{\{2\}},\{2\}} \wh{S}_{\vec{u}_{\{3\}},\{3\}}  |\psi_{1,\{1\};2,\{2\};2,\{3\}}\ra&=&
u_{\{2\},3}u_{\{3\},3} \label{spincorrunent23} \\ 
\la\psi_{1,\{1\};2,\{2\};2,\{3\}}| \wh{S}_{\vec{u}_{\{3\}},\{3\}} \wh{S}_{\vec{u}_{\{1\}},\{1\}}  |\psi_{1,\{1\};2,\{2\};2,\{3\}}\ra&=&
-u_{\{3\},3}u_{\{1\},3}  \label{spincorrunent31}
\eea
\xxx{spincorrunent12,spincorrunent23,spincorrunent31}
\yyy{OD-587 b1, OD-588 b1, OD-589 b1   \textcolor{green}{\usym{1F5F8}}  }
These are the correlations we expect to see for three unentangled spins with one spin-up and two spin-down along the $x_3$\/ axis; see Appendix~A (and set to zero the entangling coupling $\lambda$\/ that appears there.)

\section{Deutsch-Hayden representation for unentangled state}\label{SecDHrep}
\xxx{SecDHrep}

\subsection{Deutsch-Hayden transformation}\label{SecDHtransf}
\xxx{SecDHtransf}

As mentioned in Sec.~\ref{SecIntro}, the Deutsch-Hayden transformation is a unitary transformation that takes the state at $t=0$\/ to a standard state containing no information about the physical state of the system.  For quantum field theory a natural choice for such a standard state is the vacuum state, and that is the choice we make here.  We therefore wish to construct a unitary operator $\wh{V}$\/ with the property
\be
\wh{V}|\psi_{1,\{1\};2,\{2\};2,\{3\}}\ra=|0\ra. \label{Vdef}
\ee
\xxx{Vdef}
\yyy{OD-479 b3  \textcolor{green}{\usym{1F5F8}}    }

We will construct $\wh{V}$\/ sequentially as a product of three unitary operators each of which removes a particle in one region.  Define 
\ttt{START 1:49pm Sat 4/9/22}
\be
\wh{W}_{1,\{1\}}=g_1\int d^3\vec{x}d^3\vec{y} 
                            \left(\psi^*_{\{1\}}(\vec{x})\psi^*_{(1)}(\vec{y})\wh{\alpha}_{(1)}(\vec{y})\wh{\phi}_1(\vec{x})
                                   -\psi_{\{1\}}(\vec{x})\psi_{(1)}(\vec{y})\wh{\phi}^\dag_1(\vec{x})\wh{\alpha}^\dag_{(1)}(\vec{y})\right)\label{W1}
\ee
\xxx{W1}   
\yyy{OD-323 b1  \textcolor{green}{\usym{1F5F8}}    } 
with the constant $g_1$\/ real,
\be
g_1^*=g_1,\label{g1real}
\ee
\xxx{g1real  \textcolor{green}{\usym{1F5F8}}     }
\yyy{OD-323, b-1}
so $\wh{W}_{1,\{1\}}$\/ is skew-Hermitian,
\be
\wh{W}_{1,\{1\}}^\dag=-\wh{W}_{1,\{1\}}.\label{W1skew}
\ee
\xxx{W1skew}
\yyy{OD-323, b-2 \textcolor{green}{\usym{1F5F8}}   }
\ttt{STOP 2:20pm. Sat 4/9/22 --- 32 m.}

Applying the ETARs (\ref{ETAR1})-(\ref{auxETAR2}), in particular  the consequences of the fermionic nature of the field operators such as
\be
\int d^3\vec{x} d^3\vec{x}_1 \psi_{\{1\}}(\vec{x}) \psi_{\{1\}}(\vec{x}_1)\wh{\phi}^\dag_1(\vec{x})\wh{\phi}^\dag_1(\vec{x}_1)
=0,\label{symantisym}
\ee
\xxx{symantisym}
\yyy{OD-441 b-1  \textcolor{green}{\usym{1F5F8}}  }
as well as the normalization conditions (\ref{wwnormalization}), (\ref{auxnormalization}), we find the action of $\wh{W}_{1,\{1\}}$\/ on the unentangled three-particle state to be
\be
\wh{W}_{1,\{1\}}|\psi_{1,\{1\};2,\{2\};2,\{3\}}\ra=g_1|\psi_{2,\{2\};2,\{3\}}\ra,\label{W1psi122}
\ee
\xxx{W1psi122}
\yyy{OD-443 b2   \textcolor{green}{\usym{1F5F8}}  }
where
%
$$
\hspace*{-2 in}|\psi_{2,\{2\};2,\{3\}}\ra=\int  d^3\vec{x}_2 d^3\vec{x}_3 
                                                       d^3\vec{y}_2 d^3\vec{y}_3  
$$
$$
\psi_{\{2\}}(\vec{x}_2)\psi_{\{3\}}(\vec{x}_3) 
\psi_{(2)}(\vec{y}_2)\psi_{(3)}(\vec{y}_3)
$$
\be
\hspace*{2 in}\wh{\phi}_2^\dag (\vec{x}_2)\wh{\phi}_2^\dag (\vec{x}_3)
\wh{\alpha}_{(2)}^\dag (\vec{y}_2)
\wh{\alpha}_{(3)}^\dag (\vec{y}_3)|0\ra.\label{unentusualstate23}   
\ee
\xxx{unentusualstate23}  
\yyy{OD-443 b1, OD-452 b-1  \textcolor{green}{\usym{1F5F8}}    }
Application of $\wh{W}_{1,\{1\}}$\/ to $|\psi_{2,\{2\};2,\{3\}}\ra$\/ gives
\be
\wh{W}_{1,\{1\}}|\psi_{2,\{2\};2,\{3\}}\ra=-g_1|\psi_{1,\{1\};2,\{2\};2,\{3\}}\ra.\label{W1psi22}
\ee
\xxx{W1psi22}
\yyy{OD-454 b1  \textcolor{green}{\usym{1F5F8}}     }
So, if we define, for $\theta_1$\/ a real number,
\be
\wh{V}(\theta_1)_{1,\{1\}}=\exp\left(\theta_1 \wh{W}_{1,\{1\}}\right),\label{Vtheta1}
\ee
\xxx{Vtheta1}
\yyy{OD-327 b2    \textcolor{green}{\usym{1F5F8}}   }
we find, using (\ref{W1psi122}), (\ref{W1psi22}) and mathematical induction, that 
\be
\wh{V}(\theta_1)_{1,\{1\}}|\psi_{1,\{1\};2,\{2\};2,\{3\}}\ra
=\cos\left(\theta_1 g_1\right)|\psi_{1,\{1\};2,\{2\};2,\{3\}}\ra+\sin\left(\theta_1 g_1\right)|\psi_{2,\{2\};2,\{3\}}\ra.\label{Vtheta1action}
\ee
\xxx{Vtheta1action}
\yyy{OD-458 b1   \textcolor{green}{\usym{1F5F8}}    }
Choose $\theta_1$\/  and $g_1$\/ so that 
\be
\cos\left(\theta_1 g_1\right)=0,\label{cosg1thetaeq0}
\ee
\xxx{cosg1thetaeq0}
\yyy{OD-458 b-2   \textcolor{green}{\usym{1F5F8}}    }
and define
\be
s_1=\sin\left(\theta_1 g_1\right)=\pm 1.\label{s1def}
\ee
\xxx{s1def}
\yyy{OD-458 b-1, OD-459 b1, b2    \textcolor{green}{\usym{1F5F8}}       }
Defining 
\be
\wh{V}_{1,\{1\}}=\wh{V}(\theta_1)_{1,\{1\}}
\ee
where $\theta_1$\/  and $g_1$\/ satisfy (\ref{cosg1thetaeq0}), (\ref{s1def}), we obtain
\be
\wh{V}_{1,\{1\}})|\psi_{1,\{1\};2,\{2\};2,\{3\}}\ra=s_1|\psi_{2,\{2\};2,\{3\}}\ra.\label{V1psi123s1}
\ee
\xxx{V1psi123s1}
\yyy{OD-459 b3  \textcolor{green}{\usym{1F5F8}}     }

Similarly, define
\be
\wh{W}_{2,\{2\}}=g_2\int d^3\vec{x}d^3\vec{y} 
                            \left(\psi^*_{\{2\}}(\vec{x})\psi^*_{(2)}(\vec{y})\wh{\alpha}_{(2)}(\vec{y})\wh{\phi}_2(\vec{x})
                                   -\psi_{\{2\}}(\vec{x})\psi_{(2)}(\vec{y})\wh{\phi}^\dag_2(\vec{x})\wh{\alpha}^\dag_{(2)}(\vec{y})\right).\label{W2}
\ee
\xxx{W2}   
\yyy{OD-325 b2  \textcolor{green}{\usym{1F5F8}}   } 
Because both regions 2 and 3 have  particles of the same spin we must make use of the  the WSW conditions  (\ref{wswconditions})
 in addition to the ETARs (\ref{ETAR1})-(\ref{auxETAR2}) and normalization conditions (\ref{wwnormalization}), (\ref{auxnormalization}) to obtain
\bea
\wh{W}_{2,\{2\}}|\psi_{2,\{2\};2,\{3\}}\ra& = & -g_2 |\psi_{2,\{3\}}\ra, \label{W2psi22}\\
\wh{W}_{2,\{2\}}|\psi_{2,\{3\}}\ra& = & g_2 |\psi_{2,\{2\};2,\{3\}}\ra,   \label{W2psi2}
\eea
\xxx{W2psi22,W2psi2}
\yyy{OD-461 b-1, OD-467 b-1 \textcolor{green}{\usym{1F5F8}}    }
where
\be
|\psi_{2,\{3\}}\ra=\int  d^3\vec{x}_3 
                                                       d^3\vec{y}_3  
\psi_{\{3\}}(\vec{x}_3) 
\psi_{(3)}(\vec{y}_3)
{\phi}_2^\dag (\vec{x}_3)
\wh{\alpha}_{(3)}^\dag (\vec{y}_3)|0\ra.\label{unentusualstate23}   
\ee
\xxx{unentusualstate3}  
\yyy{OD-461 b-2   \textcolor{green}{\usym{1F5F8}}  }
\tstart{1:15 Tues Apr. 12, 2022} Exponentiating  $\wh{W}_{2,\{2\}}$\/,
\be
\wh{V}(\theta_2)_{2,\{2\}}=\exp\left(\theta_2 \wh{W}_{2,\{2\}}\right),\label{Vtheta2}
\ee
\xxx{Vtheta2}
\yyy{OD-327 b2, OD-461 b2   \textcolor{green}{\usym{1F5F8}} }
\tstartb{5:24pm, Thurs June 30, 2022}
whence
\be
\wh{V}(\theta_2)_{2,\{2\}}|\psi_{2,\{2\};2,\{3\}}\ra
=\cos\left(\theta_2 g_2\right)|\psi_{2,\{2\};2,\{3\}}\ra-\sin\left(\theta_1 g_1\right)|\psi_{2,\{3\}}\ra.\label{Vtheta2action}
\ee
\xxx{Vtheta2action}
\yyy{OD-470 b-1   \textcolor{green}{\usym{1F5F8}}    }
Choosing $\theta_2$\/  and $g_2$\/ so that 
\be
\cos\left(\theta_2 g_2\right)=0 \label{cosg2thetaeq0}
\ee
\xxx{cosg2thetaeq0}
\yyy{OD-471 b2, b3\textcolor{green}{\usym{1F5F8}} }
and defining
\be
s_2=\sin\left(\theta_2 g_2\right)=\pm 1,\label{s2def}
\ee
\xxx{s2def}
\yyy{OD-471 b4, b5   \textcolor{green}{\usym{1F5F8}}   }
it follows that
\be
\wh{V}_{2,\{2\}})|\psi_{2,\{2\};2,\{3\}}\ra=-s_2|\psi_{2,\{3\}}\ra,\label{V2psi23s2}
\ee
\xxx{V2psi23s2}
\yyy{OD-471 b-1   \textcolor{green}{\usym{1F5F8}}    }
where \tstop{1:50pm, Tues. Apr 12, 2022 -- 35m.}
\tstart{2:13pm, Tues. Apr 12, 2022}
\be
\wh{V}_{2,\{2\}}=\wh{V}(\theta_2)_{2,\{2\}}
\ee
with $\theta_2$\/  and $g_2$\/ satisfying (\ref{cosg2thetaeq0}), (\ref{s2def}).

Finally, to remove the spin-down particle in region 3, define
\be
\wh{W}_{2,\{3\}}=g_3\int d^3\vec{x}d^3\vec{y} 
                            \left(\psi^*_{\{3\}}(\vec{x})\psi^*_{(3)}(\vec{y})\wh{\alpha}_{(3)}(\vec{y})\wh{\phi}_2(\vec{x})
                                   -\psi_{\{3\}}(\vec{x})\psi_{(3)}(\vec{y})\wh{\phi}^\dag_2(\vec{x})\wh{\alpha}^\dag_{(3)}(\vec{y})\right),\label{W3}
\ee
\xxx{W3}   
\yyy{OD-472 b1  \textcolor{green}{\usym{1F5F8}}     } 
satisfying
\bea
\wh{W}_{2,\{3\}}|\psi_{2,\{3\}}\ra&=&g_3|0\ra\label{W3psi2}\\
\wh{W}_{2,\{3\}}|0\ra&=&-g_3|\psi_{2,\{3\}}\ra\label{W3vac},
\eea
\xxx{W3psi2,W3vac}
\yyy{OD-474 b1, b3  \textcolor{green}{\usym{1F5F8}}  }
so 
\be
\wh{V}(\theta_3)_{2,\{3\}}=\exp\left(\theta_3 \wh{W}_{2,\{3\}}\right)\label{Vtheta3}
\ee
\xxx{Vtheta3}
\yyy{OD-477 b1   \textcolor{green}{\usym{1F5F8}}   }
satisifes
\be
\wh{V}(\theta_3)_{2,\{3\}}|\psi_{2,\{3\}}\ra
=\cos\left(\theta_3 g_3\right)|\psi_{2,\{3\}}\ra+\sin\left(\theta_3 g_3\right)|0\ra.\label{Vtheta3action}
\ee
\xxx{Vtheta3action}
\yyy{OD-477 b2  \textcolor{green}{\usym{1F5F8}}     }
Choosing $\theta_3$\/  and $g_3$\/ so that 
\be
\cos\left(\theta_3 g_3\right)=0,\label{cosg3thetaeq0}
\ee
\xxx{cosg3thetaeq0}
\yyy{OD-477 b3 \textcolor{green}{\usym{1F5F8}} }
and defining
\be
s_3=\sin\left(\theta_3 g_3\right)=\pm 1\label{s3def}
\ee
\xxx{s3def}
\yyy{OD-477 b4, b5   \textcolor{green}{\usym{1F5F8}}  }
and
\be
\wh{V}_{2,\{3\}}=\wh{V}(\theta_3)_{2,\{3\}}
\ee
with $\theta_3$\/  and $g_3$\/ satisfying (\ref{cosg3thetaeq0}), (\ref{s3def}),
we have
\be
\wh{V}_{2,\{3\}})|\psi_{2,\{3\}}\ra=s_3|0\ra.\label{V3psi3s3}
\ee
\xxx{V3psi3s3}
\yyy{OD-477 b-1  \textcolor{green}{\usym{1F5F8}}   }
\tstop{2:50pm, Tues. Apr 12, 2022 -- 37m.}\tstart{3:15pm, Tues. Apr 12, 2022}

Defining
\be
\wh{V}=\wh{V}_{2,\{3\}}\wh{V}_{2,\{2\}}\wh{V}_{1,\{1\}} \label{Vproddef}
\ee
\xxx{Vproddef}
\yyy{OD-478 b1, OD-479 b2   \textcolor{green}{\usym{1F5F8}}  }
it follows from (\ref{Vtheta1}),  (\ref{Vtheta2}), (\ref{Vtheta3}) and the skew-Hermiticity of $\wh{W}_{1,\{1\}}$\/, 
$\wh{W}_{2,\{2\}}$\/ and  $\wh{W}_{2,\{3\}}$\/ that $\wh{V}$\/ is unitary, 
\xxx{OD-480 b3   \textcolor{green}{\usym{1F5F8}}    }
and from (\ref{V1psi123s1}), (\ref{V2psi23s2}) and (\ref{V3psi3s3}) that \tstop{3:24pm, Tues. Apr 12, 2022 -- 9m.}\tstart{3:33pm, Tues. Apr 12, 2022 }
\be
\wh{V}|\psi_{1,\{1\}; 2,\{2\};2,\{3\}}\ra=-s_1 s_2 s_3|0\ra.\label{Vaction0}
\ee
\xxx{Vaction0}
\yyy{OD-478 b2   \textcolor{green}{\usym{1F5F8}}    }
So, provided the $\theta$\/'s and $g$\/'s are chosen so that 
\be
s_1 s_2 s_3=-1, \label{sprod}
\ee
\xxx{sprod}
\yyy{OD-478 b-1   \textcolor{green}{\usym{1F5F8}}     }
$\wh{V}$\/ implements the  Deutsch-Hayden transformation (\ref{Vdef}).

\subsection{Unentangled operators in the Deutsch-Hayden representation}\label{SecUnentDHops}
\xxx{SecUnentDHops}

\tstop{4:18pm, Tues. Apr 12, 2022 -- 45m. }
\tstart{5:14pm, Tues. Apr 12, 2022 }

Since the state transforms according to (\ref{Vdef}), the physical operators are transformed to  the Deutsch-Hayden representation by the relations
\tstop{5:15pm, Tues. Apr 12, 2022 -- 11m. }\tstart{12:29pm, Apr. 13, 2022}
\be
\wh{\phi}_{i ,DH}(\vec{x})=\wh{V}\wh{\phi}_i(\vec{x})\wh{V}^\dag,\hspace*{.5cm}i=1,2.\label{phiDH}
\ee
\xxx{phiDH}
\yyy{OD-482 b3  \textcolor{green}{\usym{1F5F8}} }
To explicitly relate the operators (\ref{phiDH}) in the Deutsch-Hayden representation to those in the usual representation, we make use of 
 the formula \cite[p. 222]{Veltman94} 
 \tstop{1:05pm, Wed. Apr 13, 2022  -- 36m.}
\tstart{7:14pm, Wed. Apr. 13, 2022}\tstop{~7:20pm, Wed. Apr. 13, 2022 -- ~6m.}\tstart{7:27 pm, Wed. Apr. 13, 2022}
\be
e^{-y\wh{F}}\wh{G}e^{y\wh{F}}=\wh{G}+y\left[\wh{G},\wh{F}\right]
                                                             +\frac{y^2}{2!}\left[\left[\wh{G},\wh{F}\right],\wh{F}\right]
                                                             +\frac{y^3}{3!}\left[\left[\left[\wh{G},\wh{F}\right],\wh{F}\right],\wh{F}\right]+\ldots
                                                             \label{Veltmanformula}
\ee
\xxx{Veltmanformula}
\yyy{OD-9, b1   \textcolor{green}{\usym{1F5F8}}  }
\tstop{3:58pm Thurs June 2 2022}
Using this with the ETARs (\ref{ETAR1}), (\ref{ETAR2}), (\ref{auxETAR1}), (\ref{auxETAR1b}) and (\ref{auxETAR2}), the normalization conditions (\ref{wwnormalization}), (\ref{auxnormalization}), the unitary-operator definitions (\ref{Vtheta1}), (\ref{Vtheta2}),
(\ref{Vtheta3}), (\ref{Vproddef}), the parameter conditions (\ref{cosg1thetaeq0}), (\ref{cosg2thetaeq0}), (\ref{cosg3thetaeq0}), the definitions (\ref{s1def}), (\ref{s2def}), (\ref{s3def}) and mathematical induction, (\ref{phiDH}) becomes, for $i=1$\/,
\tstop{7:57pm, Wed. Apr13, 2022 -- 30m.} \tstart{Fri. Apr. 15, 9:40pm}
\be
\wh{\phi}_{1,DH}(\vec{x})=\wh{\phi}_1(\vec{x})+\psi_{\{1\}}(\vec{x})
                                            \left[-\int d^3 \vec{x}\pr  \psi^\ast_{\{1\}}(\vec{x}\pr)\wh{\phi}_1(\vec{x} \pr)
                                            + s_1\int d^3 \vec{x}\pr  \psi_{(1)}(\vec{x}\pr)\wh{\alpha}^\dag_{(1)}(\vec{x}\pr)    \right].
                                            \label{phi1unentDH}
\ee
\xxx{phi1unentDH}
\yyy{OD-487 b-2  \textcolor{green}{\usym{1F5F8}}     }
 Using in addition the WSW conditions (\ref{wswconditions}), we obtain from (\ref{phiDH}) for $i=2$\/
\bea
\!\!\!\!\wh{\phi}_{2,DH}(\vec{x})=\wh{\phi}_2(\vec{x}) 
                                            &\!\!\!\!+\!\!\!\!& \psi_{\{2\}}(\vec{x})
                                            \left[-\int d^3 \vec{x}\pr  \psi^\ast_{\{2\}}(\vec{x}\pr)\wh{\phi}_2(\vec{x} \pr)
                                            + s_2\int d^3 \vec{x}\pr  \psi_{(2)}(\vec{x}\pr)\wh{\alpha}^\dag_{(2)}(\vec{x}\pr)    \right]
                                            \nonumber \\
                                            &\!\!\!\!+\!\!\!\!& \psi_{\{3\}}(\vec{x})
                                            \left[-\int d^3 \vec{x}\pr  \psi^\ast_{\{3\}}(\vec{x}\pr)\wh{\phi}_2(\vec{x} \pr)
                                            + s_3\int d^3 \vec{x}\pr  \psi_{(3)}(\vec{x}\pr)\wh{\alpha}^\dag_{(3)}(\vec{x}\pr)    \right] .
                                            \label{phi2unentDH}                                           
\eea
\xxx{phi2unentDH}
\yyy{OD-500 b-1  \textcolor{green}{\usym{1F5F8}}  }

\tstop{Fri. Apr. 15, 10:36pm -- 56m.}

\tstart{12:24pm Sat., Apr 16, 2022}
\tstop{1:10pm Sat., Apr 16, 2022 -- 46m.}
\tstart{2:37pm Sat., Apr 16, 2022}
\tstop{3:35pm Sat., Apr 16, 2022 -- 58m.}
\tstart{5:58pm Sat., Apr 16, 2022}
\tstop{6:14pm Sat., Apr 16, 2022 -- 16m.}
\tstart{10:00am Sun., Apr 17, 2022}

It will be useful (see Sec.~\ref{SecSpincorDH}) to have the action of the Deutsch-Hayden physical operators on the vacuum state. From
(\ref{phi1unentDH}), (\ref{phi2unentDH}) and the ETARs (\ref{ETAR1})-(\ref{auxETAR2}),
\be
\wh{\phi}_{1,DH}(\vec{x})|0\ra=s
_1\psi_{\{1\}}(\vec{x})\int d^3\vec{x}\pr \psi_{(1)}(\vec{x}\pr)\wh{\alpha}^\dag_{(1)}(\vec{x}\pr)|0\ra,\label{phi1unentDHonvac}
\ee
\xxx{phi1unentDHonvac}
\yyy{OD-501 b2   \textcolor{green}{\usym{1F5F8}}     }

\bea
\wh{\phi}_{2,DH}(\vec{x})|0\ra&\!\!\!\!\!=\!\!\!\!\!&s
_2\psi_{\{2\}}(\vec{x})\int d^3\vec{x}\pr \psi_{(2)}(\vec{x}\pr)\wh{\alpha}^\dag_{(2)}(\vec{x}\pr)|0\ra\nonumber\\
&&+\;s
_3\psi_{\{3\}}(\vec{x})\int d^3\vec{x}\pr \psi_{(3)}(\vec{x}\pr)\wh{\alpha}^\dag_{(3)}(\vec{x}\pr)|0\ra,\label{phi2unentDHonvac}
\eea
\xxx{phi2unentDHonvac}
\yyy{OD-501 b-2     \textcolor{green}{\usym{1F5F8}}     }

\tstop{10:56am Sun., Apr 17, 2022 -- 56m.}
\tstart{12:33pm Sun., Apr 17, 2022}
\tstop{12:40pm Sun., Apr 17, 2022 -- 7m.}
\tstart{9:57pm Sun., Apr 17, 2022}

\be
\wh{\phi}^\dag_{1,DH}(\vec{x})|0\ra=\wh{\phi}^\dag_1(\vec{x})|0\ra
-\psi^\ast_{\{1\}}(\vec{x})\int d^3\vec{x}\pr \psi_{\{1\}}(\vec{x}\pr)\wh{\phi}^\dag_1(\vec{x}\pr)|0\ra,\label{phi1dagunentDHonvac}
\ee
\xxx{phi1dagunentDHonvac}
\yyy{OD-501 b4     \textcolor{green}{\usym{1F5F8}}      }

\tstop{10:06pm Sun., Apr 17, 2022 -- 9m.}

\tstart{3:11pm Mon., Apr 18, 2022 }
\tstop{3:16pm Mon., Apr 18, 2022 -- 5m. }
\tstart{3:50pm Mon., Apr 18, 2022 }

\bea
\wh{\phi}^\dag_{2,DH}(\vec{x})|0\ra = \wh{\phi}^\dag_{2}(\vec{x})|0\ra 
                                & \!\!\!\!- \!\!\!\!& \psi^\ast_{\{2\}}(\vec{x})\int d^3 \vec{x}\pr \psi_{\{2\}}(\vec{x}\pr)\wh{\phi}^\dag_2(\vec{x}\pr)|0\ra
                                 \nonumber\\
                                 & \!\!\!\!-\!\!\!\!& \psi^\ast_{\{3\}}(\vec{x})\int d^3 \vec{x}\pr \psi_{\{3\}}(\vec{x}\pr)\wh{\phi}^\dag_2(\vec{x}\pr)|0\ra.
                                 \label{phi2dagunentDHonvac}
\eea
\xxx{phi2dagunentDHonvac}
\yyy{OD-502 b1     \textcolor{green}{\usym{1F5F8}}      }

\tstop{5:00pm, Mon., Apr. 18, 2022}

\tstart{10:35am, Tues., Apr. 19, 2022}

\tstop{11:06am, Tues., Apr. 19, 2022 -- 31 m.}

\tstart{8:30pm. Tues., Apr. 19, 2022}
\tstop{9:20pm. Tues., Apr. 19, 2022 -- 50m.}

\tstop{5:24 Thurs June 2 2022 -- 1h 26 m -15 min = 1h 11m}

\tstart{1:34PM Tues June 7 2022}

\subsection{Effective locality of the unentangled Deutsch-Hayden transformation}\label{SecEffloc}

\xxx{SecEffloc}

From (\ref{phi1unentDH}), (\ref{phi2unentDH}) we see that Deutsch-Hayden-transformed 
physical operators differ significantly from their values in the usual representation only at locations where quanta corresponding to those operators are present.  
By a field operator $\wh{\phi}_j(\vec{x})$\/  ``corresponding'' to particle $i$\, we mean that  $\vec{x}$\/ is in the effective support of the wavefunction $\psi_{\{i\}}(\vec{x})$\/ and the spin of particle  $i$\/ is $j$\/.

For example, examining eq.~(\ref{phi2unentDH}) and noting the factors of $\psi_{\{2\}}(\vec{x})$\/ and $\psi_{\{3\}}(\vec{x})$\/ in front of the square brackets, we see that $\wh{\phi}_{2,DH}(\vec{x})$\/ can differ significantly from $\wh{\phi}_{2}(\vec{x})$\/ only where $\vec{x}$\/ is in the effective support of either $\psi_{\{2\}}$\/ or $\psi_{\{3\}}$\/,\footnote{In this section we distinguish between a function and its value for a particular value of its argument, e.g, $\psi_{\{1\}}$\/ and $\psi_{\{1\}}(\vec{x})$\/.}  i.e., at $\vec{x}$\/  where either 
$\psi_{\{2\}}(\vec{x}) $\/ or $\psi_{\{3\}}(\vec{x})$\/ is non-negligible. (From the WSW conditions (\ref{wswconditions})  we know that at most one of these functions can be nonnegligible at any particular $\vec{x}$\/.)

Furthermore, for $\vec{x}$\/ in a given region, the difference between $\wh{\phi}_{2,DH}(\vec{x})$\/ and  $\wh{\phi}_{2}(\vec{x})$\/ only depends on the values of wavefunctions in that region.   E.g., if $\vec{x}$\/ is in the effective support of $\psi_{\{2\}}$\/, the difference between $\wh{\phi}_{2,DH}(\vec{x})$\/ and $\wh{\phi}_{2}(\vec{x})$\/   
 depends only on the wavefunction $\psi_{\{2\}}$\/, not on $\psi_{\{3\}}$\/, since $\psi_{\{2\}}$\/ but not  $\psi_{\{3\}}$\/ appears in the integral multiplying $\psi_{\{2\}}(\vec{x})$\/.

 \tstop{2:00PM Tues Jun 7 2022 -- work  26m.}
 
 \tstart{2:09PM Tues Jun 7 2022}
  
However, for the Deutsch-Hayden transformation of the physical operators to be ``effectively local'' in the sense that the difference between an operator at $\vec{x}$\/ and the Deutsch-Hayden-transformed version of that operator only depends on physical information near $\vec{x}$\/, i.e., on the values of physical wavefunctions near $\vec{x}$\/, we must impose, in addition to the WSW conditions, the requirement that the effective support of each physical wavefunction is concentrated in a single localized volume, e.g., a narrow Gaussian. Imposing this additional requirement, letting  $\vec{x}_{\{i\}}$\/ be a point about which the effective support of $\psi_{\{i\}}$\/ is localized (i.e., the only points $\vec{x}$\/ where $\psi_{\{i\}}(\vec{x})$\/ is of significant magnitude are those $\vec{x}$\/ close to $\vec{x}_{\{i\}}$\/),  and returning to the the example  of  eq.  (\ref{phi2unentDH}),  we see that if $\vec{x}$\/ is within the effective support of $\psi_{\{2\}}(\vec{x})$\/,  then the difference between $\wh{\phi}_{2,DH}(\vec{x})$\/ and $\wh{\phi}_{2}(\vec{x})$\/ only comes significantly from values of wavefunctions near $\vec{x}_{\{2\}}$\/:
The physical wavefunction $\psi_{\{2\}}(\vec{x})$\/ is only significantly different from zero for $\vec{x}\approx \vec{x}_{\{2\}}$; and  the 
integral 
$\int d^3 \vec{x}\pr  \psi^\ast_{\{2\}}(\vec{x}\pr)\wh{\phi}_2(\vec{x} \pr)$\/
that  multiplies $\psi_{\{2\}}(\vec{x})$\/ only receives significant contribution from parts of the integrand within the effective support of $\psi^\ast_{\{2\}}$\/, i.e., from the parts with $\vec{x}\pr \approx  \vec{x}_{\{2\}}$\/.
Without imposing the additional localized-volume requirement,  the difference between $\wh{\phi}_{2,DH}(\vec{x})$\/ and $\wh{\phi}_{2}(\vec{x})$\/ would depend on $\psi^\ast_{\{2\}}(\vec{x}\pr)$\/ at all $\vec{x}\pr$\/ within the effective support of $\psi_{\{2\}}$\/, not just at those $\vec{x}\pr$\/ for which $\vec{x}\pr \approx \vec{x}$\/.

It is the desire to employ a Deutsch-Hayden transformation that is effectively local that leads us to include the  auxiliary fields in the formalism, as discussed in Appendix~B.   The significance of having a  Deutsch-Hayden transformation that is effectively local is discussed in Sec.~\ref{SecDisc}.

 \tstop{2:43PM Tues Jun 7 2022 -- work 34m.} 
 
 \tstart{3:05PM Tues Jun 7 2022}
 
  \tstop{3:18PM Tues Jun 7 2022 -- work 13 m.} 
  
  \tstart{4:49PM Tues Jun 7 2022 -- break/nap 1h. 31m.}

\section{Deutsch-Hayden representation for entangled state}\label{SecEntDHrep}
\xxx{SecEntDHrep}

\subsection{Entangled operators in the Deutsch-Hayden representation from ``time-evolved'' operators in the usual representation}\label{SecTwostep}
\xxx{SecTwostep}

\tstart{3:23pm, Wed. Apr. 20, 2022} 

\tstop{??}

\tstart{8:30pm, Wed. Apr. 20, 2022}

We wish to examine field operators in a Deutsch-Hayden representation for an entangled state. The process of obtaining these field operators 
is somewhat simplified if we can express the entangled state as the ``time-evolved'' version of an unentangled state for which we already know a Deutsch-Hayden transformation.


Denote the unentangled state by $|\psi_{un}\ra$\/, and the Deutsch-Hayden transformation that maps $|\psi_{un}\ra$\/ to the vacuum state by 
$\wh{V}_{un}$\/:
\be
\wh{V}_{un}|\psi_{un}\ra=|0\ra.\label{psiunDHrelation}
\ee
\xxx{psiunDHrelation}
\yyy{OD-590     \textcolor{green}{\usym{1F5F8}}    }
Let $\wh{H}_{\utoe}$\/ be a Hamiltonian that, in the Schr\"{o}dinger picture, acting over a time interval $\Delta t$\/ takes the unentangled state $|\psi
_{un}\ra$\/ to the entangled state 
$|\psi_{en}\ra$\/;  i.e., 
\be
|\psi_{en}\ra=\exp\left(\frac{-i\wh{H}_{\utoe}  \Delta t }{ \hbar} \right)|\psi_{un}\ra. \label{stateevolve}
\ee
\xxx{stateevolve}
\yyy{OD-590     \textcolor{green}{\usym{1F5F8}}    }
Then a Deutsch-Hayden transformation for $|\psi_{en}\ra$\/, i.e., a unitary transformation that maps $|\psi_{en}\ra$\/ to the vacuum state, is 
\be
\wh{V}_{en}=\wh{V}_{un}\exp\left(\frac{i\wh{H}_{\utoe} \Delta t}{\hbar}\right) ,\label{Ven}
\ee
\xxx{Ven}
\yyy{OD-590        \textcolor{green}{\usym{1F5F8}}      }
since, from (\ref{stateevolve}) and (\ref{Ven}),
\bea
\wh{V}_{en}|\psi_{en}\ra&=&\wh{V}_{un}\exp\left(\frac{i\wh{H}_{\utoe} \Delta t}{\hbar}\right)\exp\left(\frac{-i\wh{H}_{\utoe} \Delta t}{\hbar}\right)|\psi_{un}\ra\nonumber\\
&=&\wh{V}_{un}|\psi_{un}\ra\nonumber\\
&=&|0\ra\label{Venpsien}
\eea
\xxx{Venpsien}
\yyy{OD-590      \textcolor{green}{\usym{1F5F8}}         }
using (\ref{psiunDHrelation}).

Let $\wh{\chi}$\/ denote an arbitrary Schr\"{o}dinger-picture operator in the usual representation.  Applying the 
entangled-state Deutsch-Hayden transformation (\ref{Ven}), 
the Deutsch-Hayden representation of   $\wh{\chi}$\/  is
\bea
\wh{\chi}_{DH, en}&=&\wh{V}_{en}\wh{\chi}\wh{V}^\dag_{en}\nonumber\\
&=&\wh{V}_{un}\wh{\chi}(\Delta t)\wh{V}^\dag_{un},\label{chiDHen}
\eea
\xxx{chiDHen}
\yyy{OD-590      \textcolor{green}{\usym{1F5F8}}          }
where
\be
\wh{\chi}(\Delta t)=\exp\left(\frac{i\wh{H}_{\utoe} \Delta t}{\hbar}\right)\wh{\chi}\exp\left(\frac{-i\wh{H}_{\utoe} \Delta t}{\hbar}\right)\label{chiDeltat}
\ee
\xxx{chiDeltat}
\yyy{OD-590    \textcolor{green}{\usym{1F5F8}}     }
is recognized as the usual-representation Heisenberg-picture operator, at time $t=\Delta t$\/,  in that Heisenberg picture in which states and operators are equal to their Schr\"{o}dinger-picture counterparts at  time $t=0$\/.

Note, however,  that, although we refer to $\wh{H}_{\utoe}$\/ as a Hamiltonian, this is not necessarily the physical  Hamiltonian that effects the actual Schr\"{o}dinger-picture time evolution of the state $|\psi_{un}\ra$\/. Indeed, below (Sec. \ref{SecGenent}) we will take $\wh{H}_{\utoe}$\/ to be explicitly nonlocal (eq. (\ref{H})).  We will continue to describe the transformation relating  $|\psi_{un}\ra$\/ to $|\psi_{en}\ra$\/ as ``time evolution,'' but keep in mind that the state $|\psi_{en}\ra$\/ and the operator $\wh{\chi}_{DH, en}$\/ are, respectively a Schr\"{o}dinger-picture state at time $t=0$\/ and a Schr\"{o}dinger-picture operator at time $t=0$\/, equal of course to their Heisenberg-picture counterparts at time $t=0$\/.

\tstop{9:58pm, Wed. Apr. 20, 2022 -- 1h. 28m.}

\tstart{3:06pm, Thurs. Apr. 21, 2022}

\tstop{3:58pm, Thurs. Apr. 21, 2022 -- 52m.}

Further computational simplification can be anticipated if we focus on ``slightly entangled'' states, those which are close to the unentangled states from which they arise via (\ref{stateevolve}).  That is, we approximate
\be
\exp\left(\frac{-i\wh{H}_{\utoe} \Delta t}{\hbar}\right)\approx 1-i\frac{\wh{H}_{\utoe} \Delta t}{\hbar},\label{approx}
\ee
\xxx{approx}
thereby allowing us to replace  (\ref{chiDeltat}) with
\be
\wh{\chi}(\Delta t)=\wh{\chi}+\frac{i \Delta t}{\hbar} [\wh{H}_{\utoe} , \wh{\chi}].\label{chiDeltatpert}
\ee
\yyy{chiDeltatpert}
Eq. (\ref{chiDHen}) then becomes 
\be
\wh{\chi}_{DH,en}=\wh{\chi}_{DH}+\frac{i \Delta t}{\hbar} [\wh{H}_{\utoe, DH} , \wh{\chi}_{DH}], \label{chiDHenpert}
\ee
\yyy{chiDHenpert}
where
\be
\wh{\chi}_{DH} =\wh{V}_{un}\wh{\chi}\wh{V}^\dag_{un},\label{chiDH}
\ee
\yyy{chiDH}
\be
\wh{H}_{\utoe, DH}=\wh{V}_{un}\wh{H}_{\utoe}\wh{V}^\dag_{un}.\label{HutoeDH}
\ee
\yyy{HutoeDH}

\tstart{7:25pm Fri Jun 17, 2022}
\tstop{7:55pm Jun 17, 2022 -- work 30m. }
\tstart{8:10pm Jun 17, 2022 -- break 15m. }
\tstop{9:11pm Jun 17, 2022 -- work 1h. 1m. }
\tstart{9:41pm Jun 17, 2022 -- break 30m.}

\subsection{Generation of entanglement}\label{SecGenent}
\xxx{SecGenent}

\tstart{7:54pm, Thurs. Apr. 21, 2022}

\sloppy
For a Hamiltonian 
which, in the  Schr\"{o}dinger-picture, would cause the unentangled state $|\psi_{1,\{1\};2,\{2\};2,\{3\}}\ra$\/ to evolve over a time interval $\Delta t$\/ to an entangled state $|\psi_{1,\{1\};2,\{2\};2,\{3\}}(\Delta t)\ra$\/, i.e., $\wh{H}$\/ such that
\be
|\psi_{1,\{1\};2,\{2\};2,\{3\}}(\Delta t)\ra=e^\frac{-i\wh{H} \Delta t}{\hbar}|\psi_{1,\{1\};2,\{2\};2,\{3\}}\ra,\label{tevolveunent}
\ee
\xxx{tevolveunent}
\yyy{OD-590, 594   \textcolor{green}{\usym{1F5F8}}       }
\fussy
we choose \tstop{8:08pm, Thurs. Apr. 21, 2022 -- 14m.} \tstart{9:30pm, Thurs. Apr. 21, 2022}
$$
\wh{H}= -i \lambda \int d^3\vec{x}_1  d^3\vec{x}_2  d^3\vec{z}_1  d^3\vec{z}_2
$$
$$
\left(\psi_{\{1\}}(\vec{x}_1) \psi_{\{2\}}(\vec{x}_2) \psi^\ast_{\{1\}}(\vec{z}_1) \psi^\ast_{\{2\}}(\vec{z}_2) 
\wh{\phi}\da_{2}(\vec{x}_1) \wh{\phi}\da_{1}(\vec{x}_2) \wh{\phi}_{2}(\vec{z}_2)\wh{\phi}_{1}(\vec{z}_1) \right.
$$
\be
\left.-\psi^\ast_{\{1\}}(\vec{x}_1) \psi^\ast_{\{2\}}(\vec{x}_2) \psi_{\{1\}}(\vec{z}_1) \psi_{\{2\}}(\vec{z}_2) 
\wh{\phi}\da_{1}(\vec{z}_1) \wh{\phi}\da_{2}(\vec{z}_2) \wh{\phi}_{1}(\vec{x}_2)\wh{\phi}_{2}(\vec{x}_1) \right).
\label{H}
\ee
\xxx{H}
\yyy{OD-353 b1 \textcolor{green}{\usym{1F5F8}} }
\tstartb{11:07am, Fri July 1, 2022}
The action of $\wh{H}$\/ on $|\psi_{1,\{1\};2,\{2\};2,\{3\}}\ra$\/ is, from 
(\ref{ETAR1})-(\ref{killvac}), 
(\ref{wwnormalization})-(\ref{unentusualstate}) 
and (\ref{H}), 
\be
\wh{H}|\psi_{1,\{1\};2,\{2\};2,\{3\}}\ra=-i\lambda  |\psi_{2,\{1\};1,\{2\};2,\{3\}}\ra,\label{Hpsi122}
\ee
\xxx{Hpsi122}
\yyy{OD-594  1      \textcolor{green}{\usym{1F5F8}}       }
where 
$$
\hspace*{-2 in}|\psi_{2,\{1\};1,\{2\};2,\{3\}}\ra=\int   d^3\vec{x}_1 d^3\vec{x}_2 d^3\vec{x}_3 
                                                        d^3\vec{y}_1 d^3\vec{y}_2 d^3\vec{y}_3  
$$
$$
\psi_{\{1\}}(\vec{x}_1)\psi_{\{2\}}(\vec{x}_2)\psi_{\{3\}}(\vec{x}_3) 
\psi_{(1)}(\vec{y}_1)\psi_{(2)}(\vec{y}_2)\psi_{(3)}(\vec{y}_3)
$$
\be
\hspace*{2 in}\wh{\phi}_2^\dag (\vec{x}_1)\wh{\phi}_1^\dag (\vec{x}_2)\wh{\phi}_2^\dag (\vec{x}_3)
\wh{\alpha}_{(1)}^\dag (\vec{y}_1)\wh{\alpha}_{(2)}^\dag (\vec{y}_2)
\wh{\alpha}_{(3)}^\dag (\vec{y}_3)|0\ra.\label{psi212}   
\ee
\xxx{psi212}
\yyy{OD-593 b3    \textcolor{green}{\usym{1F5F8}}  }
From this point forward we will work to first order in $\frac{\lambda \Delta t}{\hbar}$\/, so (\ref{tevolveunent}) becomes, using (\ref{Hpsi122}), 
\be
|\psi_{1,\{1\};2,\{2\};2,\{3\}}(\Delta t)\ra=|\psi_{1,\{1\},2,\{2\}.2,\{3\}}\ra-\frac{\lambda \Delta t}{\hbar}|\psi_{2,\{1\},1,\{2\}.2,\{3\}}\ra.
\label{psiusualdeltat}
\ee
\xxx{psiusualdeltat}

\tstopb{3:13pm, Tues. June 28, 2022}
\tstartb{9:06am, Wed., June 29, 2022}
\tstopb{9:47am, Wed., June 29, 2022 -- work 41m.} 

To verify that the action of $\wh{H}$\/ is what we desire, we work in the usual representation and use 
(\ref{ETAR1})-(\ref{killvac}), (\ref{wwnormalization})-(\ref{unentusualstate}), (\ref{Asquared})-(\ref{apertureint}), 
(\ref{locspinopuexpl}), (\ref{psi212}) and (\ref{psiusualdeltat}) to 
compute the spin expectation values and spin correlations in the state $|\psi_{1,\{1\};2,\{2\};2,\{3\}}(\Delta t)\ra$\/, obtaining

\bea
 \la \psi_{1,\{1\};2,\{2\};2,\{3\}}(\Delta t) | \wh{S}_{\vec{u}_{\{1\}},\{1\}}|  \psi_{1,\{1\};2,\{2\};2,\{3\}}(\Delta t)\  \ra &=&
u_{\{1\},3},  \label{spinexp1en} \\
 \la \psi_{1,\{1\};2,\{2\};2,\{3\}}(\Delta t) | \wh{S}_{\vec{u}_{\{2\}},\{2\}}|  \psi_{1,\{1\};2,\{2\};2,\{3\}}(\Delta t)\  \ra &=&
   -u_{\{2\},3},  \label{spinexp2en} \\
\la \psi_{1,\{1\};2,\{2\};2,\{3\}}(\Delta t) | \wh{S}_{\vec{u}_{\{3\}},\{3\}}|  \psi_{1,\{1\};2,\{2\};2,\{3\}}(\Delta t)\  \ra &=&
-u_{\{3\},3},   \label{spinexp3en} 
 \eea
 \xxx{spinexp1en, spinexp2en, spinexp3en}
 \yyy{OD-638 b1, OD-639 b2, OD-641 b1, OD-565 b-1  \textcolor{green}{\usym{1F5F8}}    } 

$$
\hspace*{-1in}\la\psi_{1,\{1\};2,\{2\};2,\{3\}}(\Delta t)| \wh{S}_{\vec{u}_{\{1\}},\{1\}} \wh{S}_{\vec{u}_{\{2\}},\{2\}}  |\psi_{1,\{1\};2,\{2\};2,\{3\}}(\Delta t)\ra=
$$
\be
\hspace*{2in}-\left(1-\frac{2\lambda \Delta t}{\hbar}\right)u_{\{1\},3}u_{\{2\},3} -\frac{2 \lambda \Delta t}{\hbar}\vec{u}_{\{1\}}\cdot\vec{u}_{\{2\}} ,  \label{spincorrent12} 
\ee
\bea
\la\psi_{1,\{1\};2,\{2\};2,\{3\}}(\Delta t)| \wh{S}_{\vec{u}_{\{2\}},\{2\}} \wh{S}_{\vec{u}_{\{3\}},\{3\}}  |\psi_{1,\{1\};2,\{2\};2,\{3\}}(\Delta t)\ra&=&
u_{\{2\},3}u_{\{3\},3},
\label{spincorrent23} \\ 
\la\psi_{1,\{1\};2,\{2\};2,\{3\}}(\Delta t)| \wh{S}_{\vec{u}_{\{3\}},\{3\}} \wh{S}_{\vec{u}_{\{1\}},\{1\}}  |\psi_{1,\{1\};2,\{2\};2,\{3\}}(\Delta t)\ra&=&
-u_{\{3\},3}u_{\{1\},3}.
\label{spincorrent31}
\eea
\xxx{spincorrent12,spincorrent23,spincorrent31}
\yyy{OD-613 b1, OD-623 b4, OD-623 b4   \textcolor{green}{\usym{1F5F8}}  }

These expectation values and correlations match, to $\cO(\lambda)$\/,  those calculated in the analogous first-quantized system to $\cO(\lambda^2)$\/, as shown in Appendix~A, 
eqs.~(\ref{spinexp11Q})-(\ref{spincorr311Q}).  One might have expected that the generation of entanglement between spins 1 and 2 would have decreased the correlations between those spins and spin 3, i.e., that 
the magnitudes of (\ref{spincorrent23}), (\ref{spincorrent31}) would be smaller than those of  (\ref{spincorrunent23}),  (\ref{spincorrunent31}), respectively. This expectation is correct, but the the decrease only shows up at second order in the coupling $\lambda$\ generating the entanglement; see (\ref{spincorr231Q}), (\ref{spincorr311Q})
in Appendix~A. 

\tstop{10:16pm. Apr. 21, 2022 -- 46m.}

\tstart{9:46am. Apr. 22, 2022 }

\tstop{10:33am, Apr. 22, 2022 -- 47m.}

%
\subsection{Entangled Deutsch-Hayden operators}\label{SecEntDHops}
\xxx{SecEntDHops}


We now make use of the method of Sec. \ref{SecTwostep} to obtain the entangled Deutsch-Hayden field operators, i.e. the field operators transformed from the usual representation using the Deutsch-Hayden transformation that maps the entangled state to the vacuum.  
Taking the entangling Hamiltonian  $\wh{H}_{\utoe}$\/ of Sec.~\ref{SecTwostep}  to be $\wh{H}$\/ of eq.~(\ref{H}),
the unentangled Deutsch-Hayden transformation $\wh{V}_{un}$\/ of Sec.~\ref{SecTwostep} to be $\wh{V}$\/ of eq.~(\ref{Vproddef}), 
$\wh{\chi}$\/ of Sec. \ref{SecTwostep} to be $\wh{\phi}_i(\vec{x}),$\/ $i=1,2,$\/ 
and applying (\ref{chiDHenpert})-(\ref{HutoeDH}),   we find the entangled field operators 
to be 
%
%
%
$$
\hspace*{-2.5in}\wh{\phi}_{1,DH,en}(\vec{x}) = \wh{\phi}_{1,DH}(\vec{x})+\frac{\lambda \Delta t}{\hbar}\int d^3\vec{x}_1 d^3\vec{z}_1 d^3\vec{z}_2 
$$
$$
 \left(\psi_{\{2\}}(\vec{x})\psi^\ast_{\{1\}}(\vec{z}_1)\psi^\ast_{\{2\}}(\vec{z_2})\psi_{\{1\}}(\vec{x}_1) - \psi_{\{1\}}(\vec{x})\psi^\ast_{\{1\}}(\vec{z}_2)\psi^\ast_{\{2\}}(\vec{z}_1)\psi_{\{2\}}(\vec{x}_1) \right)
$$
\be
\wh{\phi}^\dag_{2,DH}(\vec{x}_1) \wh{\phi}_{2,DH}(\vec{z}_2) \wh{\phi}_{1,DH}(\vec{z}_1),\label{phi1DHen}
\ee
\xxx{phi1DHen}
\yyy{OD-482 b5, using eq 90, chiDHenpert    \textcolor{green}{\usym{1F5F8}}    }
\tstopb{11:35am, Fri July 1 2022  -- work 28m.}
\tstartb{12:19pm, Fri July1, 2022 -- break 44m.}

$$
\hspace*{-2.5in}\wh{\phi}_{2,DH,en}(\vec{x}) = \wh{\phi}_{2,DH}(\vec{x})+\frac{\lambda \Delta t}{\hbar}\int d^3\vec{x}_1 d^3\vec{z}_1 d^3\vec{z}_2 
$$
$$
 \left(\psi_{\{1\}}(\vec{x})\psi^\ast_{\{2\}}(\vec{z}_1)\psi^\ast_{\{1\}}(\vec{z_2})\psi_{\{2\}}(\vec{x}_1) - \psi_{\{2\}}(\vec{x})\psi^\ast_{\{2\}}(\vec{z}_2)\psi^\ast_{\{1\}}(\vec{z}_1)\psi_{\{1\}}(\vec{x}_1) \right)
$$
\be
\wh{\phi}^\dag_{1,DH}(\vec{x}_1) \wh{\phi}_{1,DH}(\vec{z}_2) \wh{\phi}_{2,DH}(\vec{z}_1),\label{phi2DHen}
\ee
\xxx{phi2DHen}
\yyy{OD-482 b-1, using eq 90, chiDHenpert, \textcolor{green}{\usym{1F5F8}}  }
where the unentangled Deutsch-Hayden-representation field operators $\wh{\phi}_{1,DH}(\vec{x})$\/,   $\wh{\phi}_{2,DH}(\vec{x})$\/ are as in (\ref{phi1unentDH}), (\ref{phi2unentDH}).


\subsection{Expectation values and correlations of spins of entangled and unentangled particles calculated in the entangled Deutsch-Hayden representation}\label{SecSpincorDH}
\xxx{SecSpincorDH}

Eqs. (\ref{phi1DHen}) and (\ref{phi2DHen}) provide an answer to the question posed in the title of this paper.  In this section we verify that working in the entangled Deutsch-Hayden representation, i.e., using the entangled Deutsch-Hayden operators (\ref{phi1DHen}) and (\ref{phi2DHen}) and the Deutsch-Hayden-representation state vector $|0\ra$\/, we obtain the correct results for spin expectation values and correlations.

Transforming (\ref{locspinopuexpl}) to the entangled Deutsch-Hayden representation,
$$
  \hspace*{-3in}\wh{S}_{\vec{u}_{\{i\}},\{i\},DH,en}=\int d^3 \vec{x} \; A_{\{i\}}(\vec{x}) \nonumber 
$$
$$
 \hspace*{-1in}\left[ \rule{0 mm}{5 mm} \cos \theta_{\{i\}}  \left(\wh{\phi}^\dag _{1,DH,en}(\vec{x})\wh{\phi} _{1,DH,en}(\vec{x})
                                 -\wh{\phi}^\dag _{2,DH,en}(\vec{x})\wh{\phi} _{2,DH,en}(\vec{x})\right) \right. \nonumber 
 $$
 \be
 \hspace*{.5in}+ \left.  \rule{0 mm}{5 mm}  \sin \theta_{\{i\}} \left(e^{i\phi_{\{i\}}} \; \wh{\phi}^\dag _{2,DH,en}(\vec{x})\wh{\phi} _{1,DH,en}(\vec{x})
                                 +e^{-i\phi_{\{i\}}}\; \wh{\phi}^\dag _{1,DH,en}(\vec{x})\wh{\phi} _{2,DH,en}(\vec{x})\right) \right].  \label{locspinopuexplEN}  
 \ee
  \xxx{locspinopuexplEN}
  \yyy{OD-504 b-1  \textcolor{green}{\usym{1F5F8}}    }
 We require the action of $\wh{S}_{\vec{u}_{\{i\}},\{i\},DH,en}$\/ on $|0\ra$\/, i.e.,
 $$
  \hspace*{-3in}\wh{S}_{\vec{u}_{\{i\}},\{i\},DH,en}|0\ra=\int d^3 \vec{x} \; A_{\{i\}}(\vec{x}) \nonumber 
$$
$$
 \hspace*{-1in}\left[ \rule{0 mm}{5 mm} \cos \theta_{\{i\}}  \left(\wh{\phi}^\dag _{1,DH,en}(\vec{x})\wh{\phi} _{1,DH,en}(\vec{x})
                                 -\wh{\phi}^\dag _{2,DH,en}(\vec{x})\wh{\phi} _{2,DH,en}(\vec{x})\right) \right. \nonumber 
 $$
 \be
 \hspace*{.5in}+ \left.  \rule{0 mm}{5 mm}  \sin \theta_{\{i\}} \left(e^{i\phi_{\{i\}}} \; \wh{\phi}^\dag _{2,DH,en}(\vec{x})\wh{\phi} _{1,DH,en}(\vec{x})
                                 +e^{-i\phi_{\{i\}}}\; \wh{\phi}^\dag _{1,DH,en}(\vec{x})\wh{\phi} _{2,DH,en}(\vec{x})\right) \right]|0\ra.  \label{locspinopuexplENvacaction}  
 \ee
  \xxx{locspinopuexplENvacaction}
  \yyy{OD-505 b1 \textcolor{green}{\usym{1F5F8}}    }
  Using (\ref{phi1unentDHonvac}), (\ref{phi2unentDHonvac}), (\ref{phi1DHen}), (\ref{phi2DHen}), the ETARs (\ref{ETAR1})-(\ref{auxETAR2})  and the WSW conditions (\ref{wswconditions}), 
  $$
  \wh{\phi}_{1,DH,en}|0\ra=s_1\psi_{\{1\}}(\vec{x})\int d^3\vec{x}\pr \psi_{(1)}(\vec{x}\pr) \wh{\alpha}^\dag_{(1)}(\vec{x}\pr)|0\ra
  -s_1s_2\frac{\lambda\Delta t}{\hbar}\psi_{\{2\}}(\vec{x}) 
  $$
  \be
  \cdot \int d^3 \vec{x}_1 d^3 \vec{x}\ppr  d^3 \vec{y}\ppr \psi_{\{1\}} (\vec{x}_1)\psi_{(1)}(\vec{x}\ppr) \psi_{(2)}(\vec{y}\ppr) \wh{\phi}^\dag_2(\vec{x}_1) \wh{\alpha}^\dag_{(1)}(\vec{x}\ppr) \wh{\alpha}^\dag _{(2)}(\vec{y}\ppr) |0\ra, \label{phi1DHenonvac}
 \ee
 \xxx{phi1DHenonvac}
 \yyy{OD-507, b2   \textcolor{green}{\usym{1F5F8}}    } 
 $$
  \wh{\phi}_{2, DH, en}|0\ra=s_2\psi_{\{2\}}(\vec{x})\int d^3\vec{x}\pr \psi_{(2)}(\vec{x}\pr) \wh{\alpha}^\dag_{(2)}(\vec{x}\pr)|0\ra
  $$
  $$
  +s_3\psi_{\{3\}}(\vec{x})\int d^3\vec{x}\pr \psi_{(3)}(\vec{x}\pr) \wh{\alpha}^\dag_{(3)}(\vec{x}\pr)|0\ra
  +s_1s_2\frac{\lambda\Delta t}{\hbar}\psi_{\{1\}}(\vec{x}) 
  $$
  \be
  \cdot \int d^3 \vec{x}_1 d^3 \vec{x}\ppr  d^3 \vec{y}\ppr \psi_{\{2\}} (\vec{x}_1)\psi_{(1)}(\vec{x}\ppr) \psi_{(2)}(\vec{y}\ppr) \wh{\phi}^\dag_1(\vec{x}_1) \wh{\alpha}^\dag_{(1)}(\vec{x}\ppr) \wh{\alpha}^\dag _{(2)}(\vec{y}\ppr) |0\ra. \label{phi2DHenonvac}
 \ee
 \xxx{phi2DHenonvac}
 \yyy{OD-513, b-1   \textcolor{green}{\usym{1F5F8}}     }
 Using (\ref{phi1unentDH}),  (\ref{phi2unentDH}),  (\ref{phi1DHen}), (\ref{phi2DHen}), (\ref{phi1DHenonvac}),  (\ref{phi2DHenonvac}), the ETARs (\ref{ETAR1})-(\ref{auxETAR2}),  WSW conditions (\ref{wswconditions}), normalizations (\ref{wwnormalization}), (\ref{auxnormalization}) and  the aperture conditions (\ref{Asquared})-(\ref{apertureint}) in (\ref{locspinopuexplENvacaction}), we find
 \tstopb{12:45pm, Fri. July 1, 2022 -- work 24m. }
 \tstartb{1:17pm, Fri. July 1, 2022 -- lunch 32m.}

$$
 \wh{S}_{\vec{u}_{\{1\}},\{1\},DH,en}|0\ra=\cos\theta_{\{1\}}|0\ra+\sin \theta_{\{1\}}e^{i\phi_{\{1\}}}s_1
 \int d^3\vec{x}d^3\vec{x}\pr \psi_{\{1\}}(\vec{x}) \psi_{(1)}(\vec{x}\pr)  \wh{\phi}^\dag_2(\vec{x}) \wh{\alpha}^\dag_{(1)}(\vec{x}\pr) |0\ra
 $$
 $$
 +\frac{\lambda \Delta t}{\hbar}\left( 2 \cos \theta_{\{1\}}s_1 s_2 \int d^3\vec{z}_1 d^3\vec{z}_2  d^3\vec{x}\pr d^3\vec{y}\pr 
 \psi_{\{1\}}(\vec{z}_2) \psi_{\{2\}}(\vec{z}_1) \psi_{(1)}(\vec{x}\pr) \psi_{(2)}(\vec{y}\pr)  \right.
 $$
 $$
 \wh{\phi}_1^\dag(\vec{z}_1) \wh{\phi}_2^\dag(\vec{z}_2) \wh{\alpha}_{(1)}^\dag(\vec{x}\pr) \wh{\alpha}_{(2)}^\dag(\vec{y}\pr) |0\ra
 $$
 \be
 \left.-\sin\theta_{\{1\}}e^{-i\phi_{\{1\}}}s_2 \int d^3\vec{x}_1 d^3\vec{y} \psi_{\{2\}}(\vec{x}_1) \psi _{(2)}(\vec{y}) \wh{\phi}^\dag_1(\vec{x}_1) \wh{\alpha}^\dag_{(2)}(\vec{y}) |0\ra\right),\label{Su11DHenonvac}
 \ee
 \xxx{Su11DHenonvac}
 \yyy{OD-534, b1 \textcolor{green}{\usym{1F5F8}}     }   
 $$
 \wh{S}_{\vec{u}_{\{2\}},\{2\},DH,en}|0\ra=-\cos\theta_{\{2\}}|0\ra+\sin \theta_{\{2\}}e^{-i\phi_{\{2\}}}s_2
 \int d^3\vec{x}d^3\vec{y} \psi_{\{2\}}(\vec{x}) \psi_{(2)}(\vec{y})  \wh{\phi}^\dag_1(\vec{x}) \wh{\alpha}^\dag_{(2)}(\vec{y}) |0\ra
 $$
 $$
 -\frac{\lambda \Delta t}{\hbar}\left( 2 \cos \theta_{\{2\}}s_1 s_2 \int d^3\vec{x} d^3\vec{x}\pr_1  d^3\vec{x}\pr d^3\vec{y}\pr 
 \psi_{\{1\}}(\vec{x}\pr_1) \psi_{\{2\}}(\vec{x}) \psi_{(1)}(\vec{x}\pr) \psi_{(2)}(\vec{y}\pr)  \right.
 $$
 $$
 \wh{\phi}_1^\dag(\vec{x}) \wh{\phi}_2^\dag(\vec{x}\pr_1) \wh{\alpha}_{(1)}^\dag(\vec{x}\pr) \wh{\alpha}_{(2)}^\dag(\vec{y}\pr) |0\ra
 $$
 \be
 \left.+\sin\theta_{\{2\}}e^{i\phi_{\{2\}}}s_1\int d^3\vec{x}_1 d^3\vec{x}\ppr \psi_{\{1\}}(\vec{x}_1) \psi _{(1)}(\vec{x}\ppr) \wh{\phi}^\dag_2(\vec{x}_1) \wh{\alpha}^\dag_{(1)}(\vec{x}\ppr) |0\ra\right),\label{Su22DHenonvac}
 \ee
 \xxx{Su22DHenonvac}
 \yyy{OD-536, b1    \textcolor{green}{\usym{1F5F8}}    }
\be
 \wh{S}_{\vec{u}_{\{3\}},\{3\},DH,en}|0\ra=-\cos\theta_{\{3\}}|0\ra+\sin \theta_{\{3\}}e^{-i\phi_{\{3\}}}s_3
 \int d^3\vec{x}d^3\vec{y} \psi_{\{3\}}(\vec{x}) \psi_{(3)}(\vec{y})  \wh{\phi}^\dag_1(\vec{x}) \wh{\alpha}^\dag_{(3)}(\vec{y}) |0\ra.
 \label{Su33DHenonvac}
 \ee
 \xxx{Su33DHenonvac}
 \yyy{OD-536, b2}
 
Since creation operators acting to the left annihilate the vacuum, we obtain immediately from 
 (\ref{Su11DHenonvac})-(\ref{Su33DHenonvac}) the expectation values of spin in the three regions:
\bea
 \la0| \wh{S}_{\vec{u}_{\{1\}},\{1\},DH,en}|0\ra &=&
u_{\{1\},3},  \label{spinexp1DHen} \\
 \la0| \wh{S}_{\vec{u}_{\{2\}},\{2\},DH,en}|0\ra &=&
   -u_{\{2\},3},  \label{spinexp2DHen} \\
 \la0| \wh{S}_{\vec{u}_{\{3\}},\{3\},DH,en}|0\ra &=&
   -u_{\{3\},3}.    \label{spinexp3DHen} 
 \eea
 \yyy{spinexp1DHen, spinexp2DHen, spinexp3DHen}
From (\ref{Su11DHenonvac})-(\ref{Su33DHenonvac}),  the Hermiticity of the localized spin operators, the ETARs (\ref{ETAR1})-(\ref{auxETAR2}), the WSW conditions (\ref{wswconditions}) and the normalizations    (\ref{wwnormalization}), (\ref{auxnormalization})  we compute the spin correlations:
\bea
\hspace*{-.5in}\la0|\wh{S}_{\vec{u}_{\{1\}},{\{1\}},DH,en}\wh{S}_{\vec{u}_{\{2\}},{\{2\}},DH,en}|0\ra&=&-\left(1-\frac{2\lambda \Delta t}{\hbar}\right)u_{\{1\},3}u_{\{2\},3}
-\frac{2\lambda \Delta t}{\hbar}\vec{u}_{\{1\}}\cdot\vec{u}_{\{2\}},\label{S12corrDHen}\\
\hspace*{-.5in}\la0|\wh{S}_{\vec{u}_{\{2\}},{\{2\}},DH,en}\wh{S}_{\vec{u}_{\{3\}},{\{3\}},DH,en}|0\ra&=&u_{\{2\},3}u_{\{3\},3},\label{S23corrDHen}\\
\hspace*{-.5in}\la0|\wh{S}_{\vec{u}_{\{3\}},{\{3\}},DH,en}\wh{S}_{\vec{u}_{\{1\}},{\{1\}},DH,en}|0\ra&=&-u_{\{3\},3}u_{\{1\},3}.\label{S31corrDHen}
\eea
\xxx{S12corrDHen,S23corrDHen,S31corrDHen}
\yyy{OD-566 b1, OD-567 b2, OD-566 b-1 and Hermiticity    \textcolor{green}{\usym{1F5F8}}  }
These are in agreement with the expectation values and correlations  calculated in the usual representation, (\ref{spinexp1en})-(\ref{spincorrent31}).
 
\section{Conclusions and discussion}\label{SecDisc}
\xxx{SecDisc}

 A Deutsch-Hayden-representation quantum field operator  can encode entanglement, in the case of a small degree of entanglement, via modification to the unentangled operator corresponding to a given particle consisting of an addition of a term containing unentangled field operators corresponding to the particle\footnote{See Sec. \ref{SecEffloc} re: correspondence between operators and particles.}   with which the given particle is entangled, as in (\ref{phi1DHen}), (\ref{phi2DHen}).
 
For example, suppose  the point $\vec{x}$\/ is in region 1, the effective support of $\psi_{\{1\}}(\vec{x})$\/.  Then from (\ref{phi1DHen}) and the WSW conditions we see that the 
difference between the entangled operator  
    \tstart{11:14am Mon June 20 2022} 
$\wh{\phi}_{1,DH,en}(\vec{x})$\/ 
and the unentangled operator $\wh{\phi}_{1,DH}(\vec{x})$\/  involves unentangled operators in region 2, the effective support of $\psi_{\{2\}}(\vec{x})$\/, weighted by the region-2 wavefunction and its conjugate, specifically $\int d^3\vec{z}_1 d^3\vec{x}_1
\psi^\ast_{\{2\}}(\vec{z}_1)
\psi_{\{2\}}(\vec{x}_1)
\wh{\phi}_{2,DH}(\vec{x}_1)\wh{\phi}_{1,DH}(\vec{z}_1)$\/.  

On the other hand, we see from (\ref{phi1DHen}), (\ref{phi2DHen}) that entangled field operators corresponding to the unentangled particle, i.e., those for which $\vec{x}$\/ are in region 3, are identical to their unentangled  counterparts. The $\cO(\lambda)$\/ terms in (\ref{phi1DHen}), (\ref{phi2DHen}), involving as they do  factors of $\psi_{\{1\}}(\vec{x})$\/  and  $\psi_{\{2\}}(\vec{x})$\/, vanish by virtue of the 
WSW conditions if $\vec{x}$\/ is in the effective support of $\psi_{\{3\}}(\vec{x})$\/.


The representation of physical properties in the entangled system provided by the Deutsch-Hayden field theory described above is separable. It is indeed the case that ``spatially separated systems are characterized by separate real states of affairs''\cite{Howard1985}.  
Expectation values of spin along arbitrary directions $\vec{u}_{\{i\}}$\/ in region $i$\/  
are encoded 
in operators $\wh{S}_{\vec{u}_{\{i\}},\{i\},DH,en}$\/ (see (\ref{spinexp1DHen})-(\ref{spinexp3DHen})) which are functions of the operators $\wh{\phi}_{j,DH,en}(\vec{x})$\/ in region $i$\/  (see(\ref{locspinopuexplEN})). It is the operators $\wh{\phi}_{j,DH,en}(\vec{x})$\/ that {\em are}\/ the ``separate real  states of affairs.''  


Of course determining correlations between
spins in two regions,  as in (\ref{S12corrDHen})-(\ref{S31corrDHen}), requires use of operators in both of those regions.
But no more, since the state vector $|0\ra$\/ carries no information.  As befits a separable system, ``fixing the states of the parts 
fixes the state of the whole \ldots The whole is `just the sum of the parts' ''\cite[p. 202]{Maudlin19}. 

The unentangled field operator $\wh{\phi}_{i,DH}(\vec{x})$\/ with $\vec{x}$\/ in a given region only depends on physical information in the same given region, i.e., values of physical wavefunctions in that region (see Sec. \ref{SecEffloc}).
For the entangled field operators this is not  the case.  The entangled field operator $\wh{\phi}_{i,DH,en}(\vec{x})$\/ with $\vec{x}$\/ in region 1 is a function of information in (possibly very distant) region 2, i.e., it depends on $\psi_{\{2\}}(\vec{x})$\/; and $\wh{\phi}_{i,DH,en}(\vec{x})$\/  in region 2 similarly depends on $\psi_{\{1\}}(\vec{x})$\/ (see (\ref{phi1DHen}),  (\ref{phi2DHen})).    
   This possible dependence on distant information is not surprising, given that the entangled operators are constructed from the unentangled ones using a generator $\wh{H}$\/ (eq. (\ref{H})) that is explicitly nonlocal.  Were we to regard $\wh{H}$\/ as the actual Hamiltonian acting on the system for a time interval $\Delta t$\/, it would involve instantaneous action-at-a-distance.  

However, in the actual physical world, entanglement is produced locally, in an action-by-contact fashion, and formalisms exist to 
 model this process. In particular, a previous paper\cite{Rubin02} by the present author examining entanglement in Deutsch-Hayden field theory employs  local interactions to generate entanglement starting from unentangled operators.\footnote{ Pachos and Solano \cite{PachosSolano03} compute the generation of entanglement  between two relativistic spin-1/2 fermions in QED.  Van Leent et al. \cite{vanLeentetal21} have analyzed, and performed, an experiment in which spatially-separated atoms are entangled using photons transmitted over 33 km of optical fiber. Note that the entanglement swapping employed in this experiment is in fact local, as demonstrated by  Hewitt-Horsman and Vedral\cite{HewittHorsmanVedral07} using the Deutsch-Hayden approach for qubits.} So, by combining the techniques employed in the present paper to construct in an effectively local manner unentangled Deutsch-Hayden field operators with the formalism\cite[pp. 79-80]{Kallen72} 
used in\cite{Rubin02} for local propagation  
 of field-theoretic information and generation of entanglement, it should be possible to present a quantum-field-theoretic model of entangled systems that is explicitly both separable and effectively local.   
   Effective locality, however, will not be present  unless the Deutsch-Hayden transformation from the usual representation  is effectively local --- hence the attention to effective locality in the present paper (see Sec. \ref{SecEffloc}).\footnote{As pointed out in\cite{Rubin11}  and discussed in Appendix~B of the present paper, the Deutsch-Hayden transformation employed in\cite{Rubin02}, lacking auxiliary fields, is not effectively local.  The Deutsch-Hayden transformation employed in \cite{Rubin11} makes use of auxiliary fields and is effectively local. However the starting point is a more complicated state with distinguishable spatially-coincident {\em entangled}\/  particles as well as three observers.  While a Deutsch-Hayden transformation that correctly maps the initial-time state to the vacuum state and yields effectively local transformations of the operators is obtained for this system, closed forms for all operators in the Deutsch-Hayden representation have not been obtained 
   and the time-dependent calculations are therefore  done in the usual representation.  So, separability is not explicit. We note also that in both of these previous papers all quanta, particles as well as observers, are distinguishable, in contrast to those in the present paper.}

Arntzenius, in agreement with  Deutsch and Hayden and the present author regarding the demonstration by Deutsch and Hayden of locality and separability in quantum computational networks, comments: ``Technicalities aside, Deutsch and Hayden could equally well have taken the Heisenberg picture in
quantum field theory and used the field states at locations in spacetime in place of qubits''\cite[p. 116]{Arntzenius12}.   The ``technicalities'' are, to say the least, important, as they specify how to map physical information present in the usual representation of quantum field theory into the  
Deutsch-Hayden representation.   
Pienaar, Myers and Ralph 
conclude, as we have, that ``finding an explicit form for the unitary [operator effecting a field-theoretic Deutsch-Hayden transformation] is a nontrivial matter''\cite{Pienaaretal11}.  A nontrivial matter, but not, however, an insuperable problem, as we have shown.\footnote{Rather than computing the Deutsch-Hayden transformation, Pienaar, Myers and Ralph\cite{Pienaaretal11} take an alternative approach to incorporating information into field theory operators, generalizing a model of photon generation by parametric amplification.  Recently Tibau~Vidal, Vedral and Marletto\cite{TibauVidaletal22} have proposed a local model for fermionic quantum field theory utilizing results of Raymond-Robichaud\cite{RaymondRobichaud17,RaymondRobichaud21} and B\'{e}dard\cite{Bedard21}. }

 \tstopb{1:51pm, Fri. July 1, 2022  -- work 34m.} 
 \tstartb{2:11pm, Fri. July 1, 2022  --  break 20m.}
 \tstopb{3:21pm, Fri. July 1, 2022  -- finish comparing eqs to notes :), 1h. 10m.}

\section*{Acknowledgments}

I would like to thank Jianbin Mao, Jacob A. Rubin and Allen J. Tino for helpful discussions. 

\appendix

\renewcommand{\theequation}{A-\arabic{equation}}
  \setcounter{equation}{0}  
  
  \section*{Appendix~A.  Expectation values and correlations of first-quantized spins to second order}\label{Appendix}
  
  \xxx{Appendix}

Consider a systems of three spins or qubits, i.e., distinguishable spin-1/2 particles with no spatial degrees of freedom. The state space of the $i^{\rm th}$\/  spin, $i=1,2,3$\/,  is spanned by the kets $|1\ra_{[i]}$\/  and  $|2\ra_{[i]}$\/  which are, respectively spin-up and spin-down with respect to the $x_3$\/ axis. An unentangled state  with particle 1 spin-up and particles 2 and 3 spin-down is 
\be
|\psi_{1Q,un}\ra=|1\ra_{[1]}|2\ra_{[2]}|2\ra_{[3]}.\label{psiun}
\ee
\xxx{psiun}
\yyy{OD-541 b4     \textcolor{green}{\usym{1F5F8}}    } 
Define a Hamiltonian $\wh{H}_{1Q}$\/ that acts nontrivially only on particles 1 and 2,
\be
\wh{H}_{1Q} =-i \lambda \left(|2\ra_{[1]} |1\ra_{[2]} \la 1|_{[1]} \la 2|_{[2]}  - |1\ra_{[1]} |2\ra_{[2]} \la 2 |_{[1]} \la 1|_{[2]} \right),
\ee
 \yyy{OD-552 b3   \textcolor{green}{\usym{1F5F8}}    }
 and the entangled state $|\psi_{1Q,en}\ra$\/ that results from this Hamiltonian acting on $|\psi_{1Q,un}\ra$\/ for a time $\Delta t$\/,
 \be
 |\psi_{1Q,en}\ra=\exp\left(\frac{-i\wh{H}_{1Q}\Delta t}{\hbar}\right)|\psi_{1Q,un}\ra.\label{psien}
  \ee
  \xxx{psien}
  \yyy{OD-552 b1, OD-553 b1    \textcolor{green}{\usym{1F5F8}}    }
 To  second  order in $\frac{\lambda \Delta t}{\hbar}$\/ ,
 \be
 |\psi_{1Q,en}\ra=\left(1-\frac{\lambda ^2\Delta t^2}{2\hbar^2}\right)|\psi_{1Q,un}\ra-\frac{\lambda \Delta t}{\hbar}|\widetilde{\psi}_{1Q,un}\ra,
 \label{psien2}
 \ee
 \xxx{psien2}
 \yyy{OD-554 b3     \textcolor{green}{\usym{1F5F8}}    }
 where
 \be
|\widetilde{\psi}_{1Q,un}\ra=|2\ra_{[1]}|1\ra_{[2]}|2\ra_{[3]}.\label{psitildeun}
\ee 
\xxx{OD-541 b5}
\yyy{psitildeun     \textcolor{green}{\usym{1F5F8}}       }

To see that  $|\psi_{1Q,en}\ra$\/ is in fact entangled use  (\ref{psiun}) and (\ref{psitildeun}) to write  (\ref{psien2}) as
\be
|\psi_{1Q,en}\ra=\left(\left(1-\frac{\lambda^2\Delta t^2}{\hbar^2}\right)|1\ra_{[1]}|2\ra_{[2]}-\frac{\lambda\Delta t}{\hbar}|2\ra_{[1]}|1\ra_{[2]}\right)| 2 \ra_{[3]}\label{psien3}
\ee
\xxx{psien3}
\yyy{OD-635.1 b1, OD-635.2 b1   \textcolor{green}{\usym{1F5F8}}   }
The first factor in (\ref{psien3}) is in the form of eq. (A5) of \cite{Greenbergeretal90}, and the analysis that follows there shows that (\ref{psien3}) is entangled provided $\lambda \Delta t \neq 0$\/ (and provided as well that $1-\frac{\lambda^2 \Delta t^2}{\hbar^2} \neq 0
$\/; but the vanishing of the latter would indicate we were well outside the range of validity of perturbation theory.) 

The operator that measures the spin (in units of $\hbar/2$\/) of the $i^{\rm th}$\/ particle  along the unit vector $\vec{u}_{[i]}$\/ is 
$\vec{u}_{[i]}\cdot \widehat{\vec{\sigma}}_{[i]}$\/, where the Pauli operators for the $i^{\rm th}$\/ particle satisfy
\bea
\wh{\sigma}_{1,[i]}|1\ra_{[i]}=|2\ra_{[i]},&\hspace*{.5in}&\wh{\sigma}_{1,[i]}|2\ra_{[i]}=|1\ra_{[i]},\nonumber \\
\wh{\sigma}_{2,[i]}|1\ra_{[i]}=i|2\ra_{[i]},&\hspace*{.5in}&\wh{\sigma}_{2,[i]}|2\ra_{[i]}=-i|1\ra_{[i]},\nonumber \\
\wh{\sigma}_{3,[i]}|1\ra_{[i]}=|1\ra_{[i]},&\hspace*{.5in}&\wh{\sigma}_{3,[i]}|2\ra_{[i]}=-|2\ra_{[i]}. \label{Paulis}
\eea
\xxx{Paulis}
\yyy{OD-537, b3    \textcolor{green}{\usym{1F5F8}}      }
Using (\ref{psien3})  with (\ref{Paulis}) we obtain the spin expectation values
\bea
\la\psi_{1Q,en}|\vec{u}_{[1]}\cdot \widehat{\vec{\sigma}}_{[1]}|\psi_{1Q,en}\ra&=&\left(1-\frac{2\lambda^2 \Delta t^2}{\hbar^2}\right) u_{[1],3},
\label{spinexp11Q}\\
\la\psi_{1Q,en}|\vec{u}_{[2]}\cdot \widehat{\vec{\sigma}}_{[2]}|\psi_{1Q,en}\ra&=&-\left(1-\frac{2\lambda^2 \Delta t^2}{\hbar^2}\right) u_{[2],3},
\label{spinexp21Q}\\
\la\psi_{1Q,en}|\vec{u}_{[3]}\cdot \widehat{\vec{\sigma}}_{[3]}|\psi_{1Q,en}\ra&=&-u_{[3],3},\label{spinexp31Q}
\eea
\xxx{spinexp11Q,spinexp21Q,spinexp31Q}
\yyy{OD-627 b2, OD-629 b2, OD-631 b2   \textcolor{green}{\usym{1F5F8}}     }
and the spin correlations
\be
\la\psi_{1Q,en}|\left(\vec{u}_{[1]}\cdot\wh{\vec{\sigma}}_{[1]}\right)\left(\vec{u}_{[2]}\cdot\wh{\vec{\sigma}}_{[2]}\right)|\psi_{1Q,en}\ra
=
-\left(1-\frac{2\lambda\Delta t}{\hbar}\right)u_{[1],3}u_{[2],3}-\frac{2\lambda\Delta t}{\hbar}\vec{u}_{[1]}\cdot\vec{u}_{[2]},
\label{spincorr121Q}
\ee
\bea
\la\psi_{1Q,en}|\left(\vec{u}_{[2]}\cdot\wh{\vec{\sigma}}_{[2]}\right)\left(\vec{u}_{[3]}\cdot\wh{\vec{\sigma}}_{[3]}\right)|\psi_{1Q,en}\ra&=&
\left(1-\frac{2\lambda^2\Delta t^2}{\hbar^2}\right)u_{[2],3}u_{[3],3},
\label{spincorr231Q}\\
\la\psi_{1Q,en}|\left(\vec{u}_{[3]}\cdot\wh{\vec{\sigma}}_{[3]}\right)\left(\vec{u}_{[1]}\cdot\wh{\vec{\sigma}}_{[1]}\right)|\psi_{1Q,en}\ra&=&
-\left(1-\frac{2\lambda^2\Delta t^2}{\hbar^2}\right)u_{[3],3}u_{[1],3}.\label{spincorr311Q}
\eea
\yyy{spincorr121Q,spincorr231Q,spincorr311Q}
\yyy{OD-556 b1, OD-558 b2, OD-557 b2 \textcolor{green}{\usym{1F5F8}} }
  
  \renewcommand{\theequation}{B-\arabic{equation}}
  \setcounter{equation}{0}  
  
 \section*{Appendix~B. Deutsch-Hayden transformation without auxiliary fields}\label{AppendixB}
 
 \xxx{AppendixB}
 
 It is possible to construct Deutsch-Hayden transformations for fermionic fields without including the auxiliary fields we employ here.  
 The difficulty with all such constructions we have examined to date, however, arises in attempting to obtain effectively local transformations of the field operators.

 For example, consider a state with a single fermionic particle,
 \be
 |\psi_1\ra=\int d^3\vec{x} \;\psi(\vec{x})\wh{\phi}^\dag(\vec{x})|0\ra. \label{psi1noaux}
 \ee
 \xxx{psi1noaux}
 \yyy{b2 OD-9    \textcolor{green}{\usym{1F5F8}}    }
 Defining 
 \be
 \wh{W}_1=\int d^3 \vec{x} \left(\psi^\ast(\vec{x})\wh{\phi}(\vec{x})-\psi(\vec{x})\wh{\phi}^\dag(\vec{x}) \right) \label{W1noaux}
 \ee
 \xxx{W1noaux}
 \yyy{b1 OD-10   \textcolor{green}{\usym{1F5F8}}   }
\\\tstop{5:55PM Tues Jun 7 2022 -- work, 1h. 6 m. }
\\\tstart{1:44PM Mon July 13, 2022}\\
 and
 \be
 \wh{V}_1(\theta)=\exp(\theta\wh{W}_1)\label{V1noaux}
 \ee
 \xxx{V1noaux}
 \yyy{b1, OD-12   \textcolor{green}{\usym{1F5F8}}    }
 we find
 \be
 \wh{V}_1(\theta)|\psi_1\ra = \cos(\theta)|\psi_1\ra + \sin(\theta)|0\ra.\label{V1thetapsi1noaux} 
 \ee
 \xxx{V1thetapsi1noaux}
 \yyy{b1, ODS-14   \textcolor{green}{\usym{1F5F8}}   }
 So
 \be
 \wh{V}_1= \wh{V}_1\left(\frac{\pi}{2}\right)\label{V1noaux}
 \ee
 \xxx{V1noaux}
 \yyy{b2, OD-14    \textcolor{green}{\usym{1F5F8}}      }
 is a Deutsch-Hayden transformation for $|\psi_1\ra$\/,
 \be
 \wh{V}_1|\psi_1\ra=|0\ra.\label{V1onvacnoaux}
 \ee
 \xxx{V1onvacnoaux}
 \yyy{b3, OD-14   \textcolor{green}{\usym{1F5F8}}     }
 
 The field operator in the Deutsch-Hayden representation is
 \bea
 \wh{\phi}_{DH}(\vec{x})&=&\wh{V}_1\wh{\phi}(\vec{x})\wh{V}^\dag_1 \nonumber \\
 &=& \wh{\phi}(\vec{x}) -\frac{\pi}{2} [\wh{\phi}(\vec{x}), \wh{W}_1] + \frac{(\pi/2)^2}{2!}[[\wh{\phi}(\vec{x}), \wh{W}_1],\wh{W}_1]-\ldots .\label{phiDHnoaux}
 \eea
 \xxx{phiDHnoaux}
 \yyy{b-1, OD-19, w/ theta=pi/2    \textcolor{green}{\usym{1F5F8}}      }
\tstop{2:50pm, Mon June 13, 2022 -- work 1h 6m}
\tstart{3:36pm Mon Jun 13 -- break 46m}
But
\be
[\wh{\phi}(\vec{x}), \wh{W}_1] = 2\wh{\phi}(\vec{x})\int d^3\vec{x}\pr \left(\psi^\ast(\vec{x}\pr)\wh{\phi}(\vec{x}\pr)
-\psi(\vec{x}\pr)\wh{\phi}^\dag(\vec{x}\pr)\right)+\psi(\vec{x}).\label{commphiWnoaux}
\ee
\xxx{commphiWnoaux}
\yyy{b1, OD-20}
Even if the effective support of $\psi(\vec{x})$\/ is within a localized volume concentrated around say $\vec{x}=\vec{x}\ppr$\/, the 
integral in the first term on the right-hand side of  (\ref{commphiWnoaux}) 
will depend on $\psi(\vec{x}\ppr)$\/  
regardless of the distance between $\vec{x}$\/ and $\vec{x}\ppr$\/.  So the first term in the difference between $\wh{\phi}_{DH}(\vec{x})$\/ and  $\wh{\phi}(\vec{x})$\/  (i.e., the second term on the right-hand side of   (\ref{phiDHnoaux}))  will depend on the value of 
$\psi(\vec{x}\ppr)$\/ 
no matter how far apart
 $\vec{x}$\/ and $\vec{x}\ppr$\/  are. 
\tstop{4:15pm Mon June 13 -- work 39m.}
\tstart{9:15pm Mon June 13, 2022}
\tstart{8:48am Tues June 14, 2022}
\tstop{9:31am Tues June 14, 2022 -- work 43m.}
\tstart{10:16am Tues June 14, 2022 -- break 33m.}
\tstart{2:26pm Tues June 14, 2022}

A similar issue arises in the case of the Deutsch-Hayden transformation without auxiliary fields employed in \cite{Rubin02}, as was pointed out in  \cite{Rubin11}.  There the change in the field operator for one of two species of fermions can be a function of the wavefunction for the other species at a distant point (see \cite[eq. (151)]{Rubin02}).

On the other hand, using auxiliary fields as in \cite{Rubin11} and the present paper, we find, e.g., that
\be
[\wh{\phi}_1(\vec{x}),\wh{W}_{1,\{1\}}]=-g_1\psi_{\{1\}}(\vec{x})\int d^3 \vec{x}\pr \psi_{(1)} (\vec{x}\pr)\wh{\alpha}^\dag_{(1)}(\vec{x}\pr),\label{commutatorwaux}
\ee
\xxx{commutatorwaux}
\yyy{b2, OD-337   \textcolor{green}{\usym{1F5F8}}    }
so the change in the field operator at $\vec{x}$\/ due to this term only depends on the physical wavefunction near $\vec{x}$\/, 
and ultimately the complete Deutsch-Hayden transformation for the field operator is effectively local as discussed in Sec. \ref{SecEffloc}.

%

None of the above is meant to imply that there with certainty does not exist an effectively-local Deutsch-Hayden transformation for nonrelativistic field theory of spin-1/2 fermions that does not employ auxiliary fields.  However,  as of this writing we have not discovered one. 

%
%

\end{document}